\documentclass[10pt, twocolumn]{article}
\usepackage[dvips]{graphicx}
\begin{document}


\newcommand{\be}{\vspace{2mm} \begin{equation}}
\newcommand{\ee}{ \vspace{2mm} \end{equation}}
\newcommand{\cN}{\mathcal{N}}
\newcommand{\br}{bremsstrahlung\ }
\newcommand{\bbeta}{\mbox{\boldmath $\beta$}}
\newcommand{\Ef}{\overline{E}_{f}}

\newcommand{\nc}{\newcommand}
\nc{\lsim}{\mbox{\raisebox{-.6ex}{~$\stackrel{<}{\sim}$~}}}
\nc{\gsim}{\mbox{\raisebox{-.6ex}{~$\stackrel{>}{\sim}$~}}}



\begin{titlepage}
\pagestyle{empty}
\rightline{McGill/98-19}
\rightline{hep-ph/9810439}
\rightline{21 Oct.\ 1998}
\vskip .4in

\begin{center}
{\large{\bf Numerical Study of \\
	Hawking Radiation Photosphere Formation\\
	  around Microscopic Black Holes}}
\end{center}
\vskip .1in

\begin{center}
James M.~Cline\\
Michael Mostoslavsky\\
{\it McGill University, Montr\'eal, Qu\'ebec H3A 2T8, Canada}


G\'eraldine Servant\footnote{Present address: {\it McGill University, 
Montr\'eal, Qu\'ebec H3A 2T8, Canada {\rm and} \\
Service de physique th\'eorique du CEA Saclay, 91191 Gif sur Yvette
 c\'edex, France}}\\
{\it Ecole Normale Sup\'erieure de
 Lyon, France}
\end{center}

\centerline{ {\bf Abstract} }
\vskip 0.5truecm

Heckler has recently argued that the Hawking radiation emitted from
microscopic black holes has sufficiently strong interactions above a
certain critical temperature that it forms a photosphere, analogous to
that of the sun.  In this case, the visible radiation is much cooler
than the central temperature at the Schwarzschild radius, in contrast
to the naive expectation for the observable spectrum.  We investigate
these ideas more quantitatively by solving the Boltzmann equation using
the test particle method.  We confirm that at least two kinds of
photospheres may form: a quark-gluon plasma for black holes of mass
$M_{BH} \lsim 5\times 10^{14}$ g and an electron-positron-photon plasma
for $M_{BH} \lsim 2\times 10^{12}$ g.  The QCD photosphere extends from
the black hole horizon to a distance of 0.2--4.0 fm for $10^9$ g$\lsim
M_{BH} \lsim 5\times 10^{14}$ g, at which point quarks and gluons with
average energy of order $\Lambda_{QCD}$ hadronize. The QED photosphere
starts at a distance of approximately 700 black hole radii and
dissipates at about 400 fm, where the average energy of the emitted
electrons, positrons and photons is inversely proportional to the black
hole temperature, and significantly higher than was found by Heckler.
The consequences of these photospheres for the cosmic diffuse gamma ray
and antiproton backgrounds are discussed: bounds on the black hole 
contribution to the density of the universe are slightly weakened.

\end{titlepage} 

\section{Introduction}

It has long been known that black holes are not perfectly black, but
emit nearly black-body radiation at a temperature $T_{BH}=(8 \pi
GM)^{-1}$ due to quantum mechanical effects \cite{Hawking}(where G is
Newton's gravitational constant). Although
this Hawking radiation is negligible for astrophysically
large black holes, it becomes sufficiently hot to be visible for masses
$M \lsim 10^{15}$ GeV, corresponding to a BH that would be disappearing
today, assuming it was present at the big bang. The present density of
such BH's in the universe is in fact limited by observations of the
diffuse gamma ray background coming from their accumulated radiation
\cite{Mac:Carr} (for a review see 
\cite{Halzen}).  Such limits are calculated assuming Hawking's expression 
for the spectrum,
\be
\label{eqn:K1}
\frac {dN}{dEdt}=\frac {\sigma_s(E)E^2}{2\pi} \frac {1}{\exp(E/T_{BH}) \pm 1}
\ee
where the sign depends on whether the emitted particle is a fermion
($+$) or a boson ($-$), and $\sigma_s(E)$ is the absorption cross
section for the emitted particle, which depends on its spin $s$
\cite{Page}.  Moreover, previous estimates of the possibility of
observing individual black holes, which explode in a burst of radiation
as their masses approach to Planck mass, are also based on
(\ref{eqn:K1}), combined with calculations of the mass spectrum of
primordial black holes (PBH's) that could form during inflation
\cite{Inflation} or the QCD phase transition \cite{QCD}, and guesses as
to how they might cluster.  It is possible that extremely high energy
neutrinos from exploding PBH's will be observable in the new generation
of neutrino telescopes \cite{Halzen2}.

Recently Heckler \cite{Heck:1} revived the possibility that the
spectrum (\ref{eqn:K1}) need not hold far away from the Schwarschild
radius, $r_H=2GM$, because the radiation might interact with itself in
some region, similar to photons diffusing inside the photosphere of the
sun.  The idea had been previously dismissed \cite{Ol:Hill,Mac:Carr},
but in the framework of QED, Heckler identified bremsstrahlung and pair
production as processes which could change this negative conclusion.
Both interactions cause a small initial number of high energy particles
to fragment into many lower energy particles, giving a significant
decrease in the average particle energy.  Consequently, according
to ref.~\cite{Heck:1}, a BH with  $T=45$ GeV at $r_H$ would appear to an
observer more like a BH with $T=m_e$ (the electron mass), in terms of
average energy, but much brighter in terms of absolute luminosity.

As recognized by Heckler, it is not easy to give a highly quantitative
treatment of the problem because, unlike in the sun, the system is
never so strongly coupled as to admit the approximation of local
thermal equilibrium. The problem is that, whereas the sun has a huge
chemical potential in gravitationally bound electrons and protons which
are providing the large density, in the BH there are equal numbers of
particles and antiparticles, which are at first freely streaming away
from the horizon. The density at the horizon is $\sim T^3$ but quickly
falls like $1/r^2$ in the absence of particle production. In fact the
density changes by a large factor within the mean free path of the
particles. A fluid description of the plasma, although perhaps useful
for roughly estimating its behavior, is not self-consistent. 

It thus seems worthwhile to investigate the evolution of the Hawking 
radiation plasma more quantitatively. The proper framework for doing so
is the Boltzmann equation for the particle distributions,
\be
\label{eqn:K2}
\left(
\frac{\partial}{\partial t} + \frac {\mathbf{p}}{E} \cdot 
\mathbf{\nabla} \right) f(\mathbf{p},\mathbf{x},t)=
{\cal C} [f(\mathbf{p},\mathbf{x},t)] 
\ee
\noindent
using (\ref{eqn:K1}) as a boundary condition at $r=r_H$. In the
following section we will discuss a general method for solving the
Boltzmann equation which has been successfully used in the field of
heavy ion collisions, and the adaptations of this method which we have
made for the BH problem. Crucial for this investigation is the
collision term~${\cal C}[f]$. Section 3 focuses on the scattering cross
sections which go into ${\cal C}$. In section 4 we present our results
for the detailed properties of the QED and QCD photospheres.
The ramifications for the most relevant observable particle backgrounds, 
namely gamma rays and antiprotons, are worked out in section 5.
We summarize our results in the final section. 


\section{Test particle method for solving Boltzmann equation} 
The full Boltzmann equation is an integro-differential equation which
is difficult to solve exactly. For black holes we can make several
important simplifications. (1)~We consider nonrotating black holes so
that the distribution functions have spherical symmetry. (2)~We confine
our attention to BH's whose lifetime is still long compared to the
diffusion time of particles in the putative photosphere; thus the
distribution functions are approximately time-independent.  For
example, the lifetime of a black hole with $M= 10^{11}$ g and $T=1000$
GeV is of order $10^{5}$ s, whereas the particle diffusion time in its
photosphere is $\sim 10^{-21}$ s. We estimate that only for black holes
of mass $\lsim 10^{7}$ g will the lifetime become comparable to the
diffusion time.

Under the above assumptions, the Boltzmann equation takes the simpler
form
\be
\label{eqn:K3}
\label{BEq-r}
\frac{\partial}{\partial r} f(p,p_t,r)=\frac{1}{v_r}{\cal C} [f],
\ee
where $p=|\mathbf{p}|$, $p_t$ is the magnitude of the component of 
momentum transverse to the radial direction, and $v_r$ is the radial 
velocity. Because of the spherical symmetry there is no dependence 
on the azimuthal momentum component.

Even with these simplifications, eq.~(\ref{eqn:K3}) is still
prohibitively difficult to solve in the most naive numerical way,
namely discretizing momentum space and evolving the distribution
defined on this momentum lattice forward in radius. At each point in
momentum space a multidimensional phase space integral is required to
evaluate the collision term, which is computationally costly.
Fortunately, this problem has been already surmounted in other
situations, namely heavy ion collisions~\cite{DasGupta, Welke}. There
one wants to track the distribution functions of nucleons in the two
nuclei as they pass through each other and undergo collisions.

The idea is essentially to follow the classical evolution of each
particle, allowing for possible scatterings by using the differential
scattering cross section as a probability distribution. However, the
number of nucleons in even a heavy nucleus is so small that one must do
this many times to obtain distribution functions that are not dominated
by statistical fluctuations. Alternatively, one can obtain the required
statistics by representing each real particle by an arbitrarily large
number of test particles, which is equivalent to but simpler than
simulating the heavy ion collision many times. One must only be sure to
avoid unphysical collisions by prohibiting any two test particles that
represent the same actual particle from scattering off each other.
However, in practice, it is easier to allow scatterings to occur
between {\it all} test particles, while simultaneously reducing the
cross section by a factor $N$, equal to the ratio between the number of
test particles and the number of actual particles.

A difference between heavy ion collisions and black hole radiation is
that in the former, one is evolving the distributions in time, whereas
we want to evolve them in space. However, our version of the Boltzmann
equation is mathematically equivalent to one with time evolution but
spatial homogeneity,
\be
\label{eqn:K4} 
\frac{\partial}{\partial t} f(\mathbf{p},t)={\cal C}'[f],
\ee
where $f(\mathbf{p},t)$ is known at some initial time. Since this 
equation is evidently amenable to solution by the test particle method, 
and eq.~(\ref{eqn:K3}) is equivalent to it just by renaming $t \rightarrow 
r$ and ${\cal C}'[f] \rightarrow {\cal C}[f]/v_r$,
we can also apply the same method to solve (\ref{eqn:K3}).

\subsection{Free evolution}

Let us first consider how the method works in the absence of collisions, 
so the particles are propagating freely. At the initial surface of the 
horizon, it is assumed that the radiation is distributed according to 
eq.~(\ref{eqn:K1}) in all directions not pointing back below the horizon, 
that is, for polar angles with $\cos\theta > 0$. However, at a larger 
radius $r$, if there have been no collisions then particles can only have 
come from a cone pointing back  to and subtended by the horizon. The 
directions of the possible momenta of the particles are restricted to 
lie in the outward half of this cone, whose opening angle is given by
$\sin\theta=r_H/r$. As $r$ increases, the momenta become 
increasingly concentrated in the radial direction.

\begin{figure}[h!]
\centering
\includegraphics[width=0.45\textwidth]{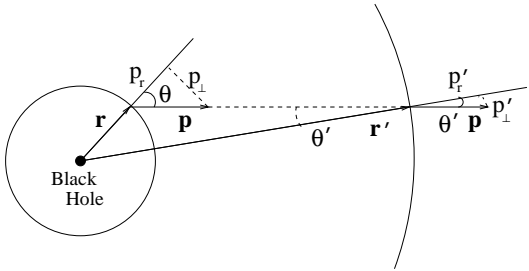}
\caption{Momentum in free radial evolution}
\label{fig:free_evolution}
\end{figure}

Numerically the evolution in the noninteracting case is thus trivial. A
particle whose momentum is $\mathbf{p}$ at $r_H$ will have the same
momentum at $r'> r$. However, the {\it coordinates} of the momenta will
change: for example, if $\mathbf{p}$ is purely transverse at $r$, it
will develop a radial component at $r'$. Let $\theta$ be the polar
angle from the radial direction at $r$ and $\theta '$ that at $r'$ (see
Fig.~\ref{fig:free_evolution}). Then using the law of sines,
\be
\label{eqn:K5}
\sin\theta'=\frac{r}{r'}\sin\theta.
\ee
\protect

For very large $r'$ the momentum becomes essentially radial, $\theta'  
\cong 0$, and a momentum that is initially radial remains so. The 
distribution function will be given approximately by
\be
\label{eqn:K6} 
 f(\mathbf{p'},r') \cong 
 \left\{ 
  \begin{array}{ll}
    f(\mathbf{p},r_H), & 0 \le \sin\theta' \le r_H/r \\
                         \\
    0, & \sin\theta' > r_H/r
  \end{array}
 \right.,
\ee
where the new radial and perpendicular components are related to the old 
ones by
\be
\label{eqn:K7} 
\label{eq:propagation}
p'_{\bot}=\frac{r_H}{r} p_{\bot} \quad \mbox{and} \quad 
p'_r=\sqrt{p^2-{p'_{\bot}}^2}.
\ee
\noindent
If $f$ is isotropic at the horizon, then for large distances one finds 
that this gives a density decreasing like $1/r^2$, as required by 
conservation of particle flux:
\be
\label{eqn:K8} 
n(r)=\int \frac{d^{\,3}p}{(2\pi)^3} f(\mathbf{p},r)
={\left(\frac{r_H}{r}\right)}^2 n(r_H).
\ee
Since the particles are moving on straight lines, the step size for 
evolving the distributions is irrelevant in the noninteracting 
case.

\subsection{Including interactions}

To account for the interactions of the particles with each other one
must first choose an appropriate step size $\Delta r$ for evolving the
distribution. The natural choice is the mean free path, $\lambda$, or
some fraction thereof. If the interaction cross section is $\sigma$,
then we can define
\be
\label{eqn:K9} 
\lambda^{-1}(r)=n(r)\langle \sigma v_{rel} \rangle,
\ee
where $n$ is the density and $v_{rel}$ is the relative velocity of the
two interacting particles (to be discussed in more detail below, in
section 3). However, the time-independent Boltzmann equation we are
solving is mathematically equivalent to the usual time-dependent
version only after dividing by the radial velocity. Thus our collision
term is related to the usual one by a factor of $1/v_r$ , which
modifies the definition of the mean free path to
\be
\label{eqn:K14}
\label{eq:meanfreepath}
\lambda^{-1}(r)=n(r)\left\langle \sigma {v_{rel}\over v_r} \right\rangle.
\ee
We take $v_r$ in this formula to be the minimum of the radial velocities 
of the two interacting particles. In practice this factor is not very 
important because the bulk flow of the particles away from the black 
hole tends to cause $v_r$ to approach unity.

The average is taken using the distribution functions for the 
incoming particles,
\be 
\label{eqn:K10} 
\left\langle \sigma {v_{rel}\over v_r} \right\rangle = 
\int \frac {d^{\,3}p_{1} d^{\,3}p_{2}}
{(2\pi)^6} 
f(\mathbf{p_1},r) f(\mathbf{p_2},r) \sigma {v_{rel}\over v_r}.
\ee 
In the test particle method, there is no need to do any integrals {\it 
per se}. Since the ensemble of test particles is already
distributed according to $f$, $\langle \sigma v_{rel}/v_r\rangle$ is 
simply given by an unweighted average of $ \sigma v_{rel}/v_r$ over the 
ensemble. We randomly choose pairs of particles to perform this average.



If $\lambda(r)$ was constant, one could choose the step size for the 
evolution to be simply $\Delta r=\lambda$. Since $\lambda$ is the average 
distance particles go between collisions, the interactions could be 
simulated by demanding that over the distance $\Delta r=\lambda$, each 
particle participates in a single collision. However, in the black hole 
problem $\lambda(r)$ can increase significantly on a distance scale of  
$\lambda$ because the density is decreasing like $1/r^2$. To deal with 
this, one must choose the step size to be smaller, 
\be
\label{eqn:K12} 
\Delta r\ll \frac{\lambda}{d\lambda/dr}.
\ee
Then $\lambda$, and hence the interaction rate, is guaranteed to be 
approximately constant over $\Delta r$.
Over this distance, only a fraction $F$ of all the particles will undergo 
scattering,
\be
F=\frac{\Delta r}{\lambda}.
\ee

The general procedure for including interactions is therefore clear:
(1)~At each $r$, compute $\lambda$ using the known distributions
$f(\mathbf{p},r)$. (2)~Given $\lambda$, choose a step size $\Delta r$
in accordance with (\ref{eqn:K12}). (3)~Choose a subset of particles
from the ensemble such that a fraction $F=\Delta r/\lambda$ of the
total ensemble will undergo collisions and change their energies and
momenta accordingly.  To account for the $1/v_r$ factor in the
effective cross section, we arrange the ensemble in order of increasing
$v_r$ and let that fraction of particles $F$ with the smallest $v_r$
participate in the interactions.  After these steps are carried out, the
distribution function is known at $r+\Delta r$ and the procedure can be
repeated.


%
%

    The particle density $n(r)$ which goes into~(\ref{eq:meanfreepath}) is 
calculated analytically and therefore does not depend on the number of 
the test particles. To find the density $n(r)$ we first compute what it 
would be in the absence of particle production: 
\be
n_0(r)=n_h \frac{r_h^2}{r^2},
\label{n0}
\ee
where the subscript 0 means that this density is before the particle
creation process is taken into account.  The radius of the horizon is
\mbox{$r_h=1/(4\pi T_{BH})$}.  To find the density at the horizon, we
note that if the BH absorption cross section $\sigma(E)$ was a
constant, then the BH would be a perfect black body, and the density of
radiation would be thermal.  However $\sigma(E)$ vanishes as $E\to 0$
and only reaches its geometric optics value of $\sigma_0 = 27 \pi G^2
M^2$ in the limit as $E\to\infty$.  Therefore $n_h$ is reduced from its
thermal value by a factor of $\Gamma_s \equiv \int d^{\,3} p\,
\sigma(E) f_\pm(p) / \int d^{\,3} p\, \sigma_0 f_\pm(p)$. This has been
computed in ref.~\cite{MacWeb} to be
\be
 \Gamma_s = \left\{ \begin{array}{ll}
                      \frac{56.7}{27\pi}, & \mbox{electrons}\\
                        \\
                      \frac{20.4}{27\pi}, & \mbox{photons.}
                     \end{array}
             \right.
\label{Gamma_s}
\ee
Then the density at the horizon is 
\be
n_h = \left\{ \begin{array}{ll}
        \frac{3}{2\pi^2} \Gamma_{\!f} \, \zeta(3)\,T_{BH}^3, & 
	\mbox{electrons or positrons}\\
                        \\
      \frac{2}{\pi^2} \Gamma_{b} \, \zeta(3)\,T_{BH}^3, & 
	\mbox{photons,}
             \end{array}
      \right.
\ee
where $\zeta(3)=1.20206$ (Rieman zeta function). 

To account for particle creation we use the test particles to find the
fraction of new fermions and bosons created at each step. Let
$N_{f[b]}(r)$ be the number of electrons [photons] in the shell
of width $\Delta r$ at radius $r$. We will define $P_{f[b]}(r)$ as:
\be
P_{f[b]}(r)=\frac{N_{f[b]}(r)}{N_{f[b]}(r_h)}.
\label{Ratio}
\ee
Then, using (\ref{n0}) through (\ref{Ratio}) we obtain
\be
n(r) = \frac{\zeta(3)}{\pi^2 {(4\pi)}^2} \frac{T_{BH}}{r^2}
       \left\{ \begin{array}{ll}
        {3\over 2} \Gamma_{\!f} \,P_f(r),  &  e^+ \mbox{\ or\ } e^-\\
                        \\
         2 \Gamma_{b} \, P_{b}(r), & \gamma.
               \end{array}
      \right.
\label{eq:density}
\ee
In this way, particle densities can be computed at any step by keeping
track of the relative increase in particle number, $P_{f[b]}$.  Later,
we shall also refer to the ratio of all particles at $r$ versus at
the horizon,
\begin{eqnarray}
	P(r) &=& {n_b(r) + n_f(r) \over n_b(r_h) + n_f(r_h)}\nonumber\\
	     &=& {P_b(r) + 4.17 P_f(r) \over 1 + 4.17}.
\label{eq:pprod}
\end{eqnarray}
The factor 4.17 comes from computing $3\Gamma_f/2\Gamma_b$.

To generalize the previous results to quarks and gluons is
straightforward: one must multiply the photon density by a factor of
$8$ to convert to gluons, and the electron density by a factor
$3 n_f$ to get $n_f$ flavors of quarks.

A final issue concerns the number of test particles used to represent
the ensemble, versus the actual number of particles coming from the
BH.  For example, within the first radial increment $\Delta r$ near the
horizon, the actual number of particles \mbox{$\Delta N = 4 \pi r^2_H
\Delta r\, n(r_H)$} might be too small a number to generate a
smooth distribution function.  We would prefer to
represent these particles with some much larger number $N_t$ of test
particles. In the application to heavy ion collisions it is necessary
to reduce the cross section by the factor $\Delta N/N_t$ to avoid
overestimating the number of collisions. This is because in the latter
situation, individual nucleon  positions are kept track of, and two
particles are only allowed to collide if they come within a distance
$b_c= \sqrt{\sigma/\pi}$ of each other. If the number of particles is
artificially increased while proportionally decreasing the cross
section, the physically meaningful mean free path will remain constant.
In our case, since we do not follow the spatial trajectories of
particles, but instead compute the physical mean free path, it is
consistent to allow {\it all} the test particles to interact over a
distance $\lambda$, regardless of how large the ratio $N_t/ \Delta N$
is. Thus there is no need to reduce $\sigma$ proportionally to the
number of test particles in the present problem.

\section{Scattering cross section}

\begin{figure}[!h]
\centering
a)\includegraphics[width=0.1\textwidth,height=0.06\textheight]
{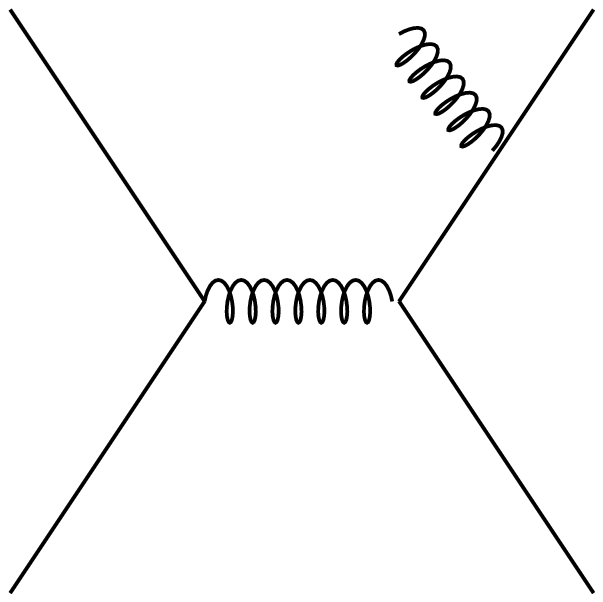} 
\hspace{1.2 cm}
b)\includegraphics[width=0.1\textwidth,height=0.06\textheight]
{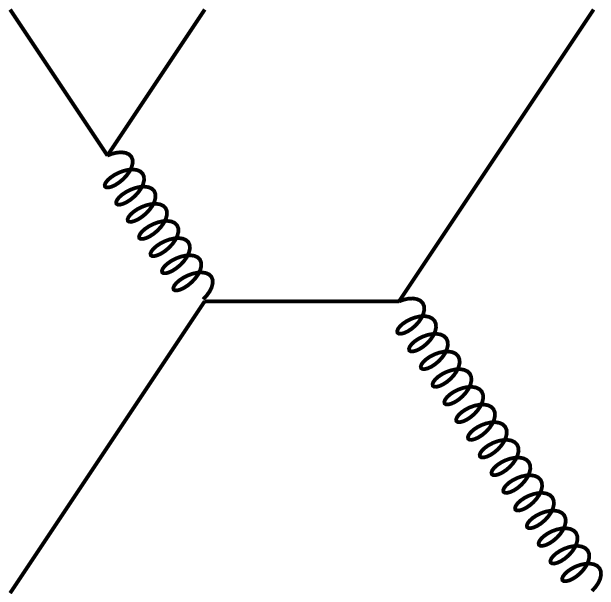}
\caption{Feynman diagrams for the dominant contributions to (a) \br; (b)
pair production.}
\label{fig:relevant}
\end{figure}

The most important processes contributing to photosphere formation around
the BH are \br ($ee\to ee\gamma$) and photon-electron pair production ($e
\gamma \to e e^- e^+ $), whose dominant Feynman diagrams are shown in
Fig.~\ref{fig:relevant}. 
\vspace{0.3cm}
\protect
\vspace{-0.2cm}
Both have cross sections of $O(\alpha^3)$, which at first sight might
make them seem less important than the leading $O(\alpha^2)$ processes
like Compton scattering.  However the latter are elastic and change the
distributions only by randomizing the momenta, whereas the former are
inelastic and increase the number of particles while reducing their
average energies.  Thus the elastic scattering processes appear to be
less relevant for determining the properties of the photosphere even
though they are faster than the inelastic ones.  Before going on to a
detailed description of brehmsstrahlung, let us examine this issue more
carefully.

\subsection{Elastic scattering processes}

One could argue that elastic interactions like Compton scattering
($e\gamma\to e\gamma$) or electron-electron scattering might change the
size of the photosphere somewhat, but they will not affect the most
important observable property, which is the energy spectrum of emitted
particles.  It is the $2\to 3$ body interactions like \br which
principally distinguish the photosphere from freely streaming particles. 
However there is one sense in which elastic scatterings may be important: 
by randomizing the particle momenta, they might postpone the tendency for
the relative velocities of particles to approach zero, due to the bulk,
outward radial motion.  This in turn could keep the \br mean free path,
eq.~(\ref{eqn:K14}), small out to larger radii, enlarging the size of the
photosphere.  

Suppose the mean free path for elastic processes is given by
$\lambda_e(r)$ at a distance $r$ from the black hole.  Their
momentum-randomizing effects will be important for the photosphere
only in regions where $\lambda_e \lsim r$.  This is because the length
scale over which random momenta become increasingly radial at a given
distance $r$ is $r$ itself.  If $\lambda_e(r)$ exceeds $r$, then
scatterings start to lose in the competition against geometry.  

The relativistic limit of the Compton cross section, in the center
of mass frame, is 
\be
	\sigma_c = {2\pi\alpha^2\over m_e E}\ln{E\over m_e}.
\ee
We can then estimate $\lambda_e$ as
\be
 \lambda_e^{-1}(r) \cong \sigma_c n(r) = \sigma_c n(r_h) P(r),
\ee
with the particle density given by (\ref{eq:density}) and the particle
production factor by (\ref{eq:pprod}).  Below, we will show that when 
elastic scattering is ignored, the QED photosphere ends at a distance of
$r_f \sim m_e^{-1}$, and for large $T_{BH}$, $P(r_f) \sim b T_{BH}^2$
with $b = 4.5\times 10^{-5}$ GeV$^{-2}$.  Furthermore, the average
particle energy at the horizon will be given by its value at the
horizon, $\sim 3T_{BH}$, divided by P(r).  To see whether 
our neglect of elastic scattering is consistent, we should compute 
$r_f/\lambda_c \cong (m_e \lambda_c)^{-1}$, and ask whether it ever
exceeds $O(1)$.  Using the above results we get
\begin{eqnarray}
	{1\over m_e \lambda_c} \cong {\sigma_c\over m_e}
	\left( n(r_h) {r_h^2\over r_f^2} P(r_f) \right)\nonumber\\
	 = 4.3 \left({T_{BH}\over 10 \hbox{\ TeV}}\right)^4
	\left(9.5 - \ln\left({T_{BH}\over 10 \hbox{\ TeV}}\right)\right)
\end{eqnarray}
Therefore the effects of elastic scattering should only become important
when the BH temperature starts to exceed $\sim 5$ TeV.  We will discuss
our numerical investigation of this regime below, although most of our
work focuses on BH temperatures below 1 TeV.

\subsection{Bremsstrahlung cross section}

The relativistic differential cross sec\-tion for bremsstrahlung in 
the center of mass frame is \cite{JR,Haug}
\begin{eqnarray}
\frac{d\sigma(\omega)}{d\omega} \approx \frac{8\alpha r_e^2}{E\omega}
\left(\frac{4}{3}(E-\omega)+\frac{\omega^2}{E} \right) \nonumber \\ \times
\left( \ln \left[\frac{4E^2(E-\omega)}{m_e^2\omega} \right] - \frac{1}{2} 
\right),
\label{eq:dsigma}
\end{eqnarray}
where $\hbar=c=1$, $r_e=\alpha/m_e$, $E$ is the initial energy of each 
electron, and $\omega$ is the energy of the emitted photon. It diverges 
for $\omega \rightarrow 0$, but higher order corrections essentially 
impose an infrared cutoff \cite{Heck:1}.

The form of Eq.~(\ref{eq:dsigma}) implies that the probability to emit
a photon diverges as its energy goes to zero.  On the other hand,
emission of a zero-energy photon has no effect on the electron which
emits it. A convenient way of rendering the cross section finite, while
at the same time accounting for the photons which carry away
significant energy, is to use the energy-averaged total cross section
\cite{Heck:1,JR}
\be
\bar\sigma=\int \frac{\omega}{E} \left(\frac{d\sigma}{d\omega} \right) 
d\omega 
\approx 8\alpha r_e^2 \ln \frac{2E}{m_e}.
\label{eq:sigma}
\ee
The cross section for photon-electron pair production shows the same
functional dependence in the extreme relativistic limit and we
therefore use the same estimate (\ref{eq:sigma}) for both
interactions.

%

An improvement on the present treatment would be to use the actual
differential bremsstrahlung cross section as a distribution which would
produce low energy photons with higher probability.  One would have to
impose an infrared cutoff on the emitted photon energy and show that no
meaningful physical properties of the photosphere depend on this
cutoff.  Here we have taken the simpler approach of approximating
$d\sigma/d\Omega = \bar\sigma/4\pi$, and choosing the energies and
directions of the final state particles at random in the center of mass
system, subject to the constraint of conservation of four-momentum.

\subsection{Thermal Mass}
Bremsstrahlung proceses do not occur in vacuum but in a background 
plasma of almost radially propagating particles. These background
particles suppress the \br cross section if they are sufficiently
dense (the LPM effect \cite{LPM}).
A simplified means of accounting for this is to replace the vacuum
electron mass, $m_{e}$, by its thermal value,
\be
m_{th}=\sqrt{m_{e}^2+m_{p}^2(T)}.
\label{eq:thmass}
\ee
This procedure, although not exact, gives the correct position of the
pole of the electron propagator in the energy-momentum plane to an
accuracy of 10\% \cite{Weldon}.
In a gauge theory with coupling constant $g$, in a thermal background,
the plasma mass is given by $m_p=gT\sqrt{C(R)/8}$, where $C(R)$ is the
quadratic Casimir invariant for the fermion transforming in the
representation $R$ of the gauge group. However, even if the
background is not in thermal equilibrium, we find by computing the 
electron self-energy in the plasma that the formula for
$m_p$ which was originally derived in the thermal case is still correct:
\be
m_{p}^2=g^2 C(R)\int {d^{\,3}p \over (2\pi)^3 p}\left(f_f(p)+
                                         f_b(p) \right),
\ee
where $f_{f(b)}(p)$ is the distribution function for a single
polarization of the background fermions (gauge bosons),  ({\it i.e.,
}$f_{f(b)}(p)  = (e^{E/T}\pm 1)^{-1}$ in the thermal case).  For a
collection of test particles which represents these distribution
functions, the integral may be approximated by taking the average of
$1/p$ over all electrons and photons, respectively (quarks and gluons
in the case of QCD).  Counting polarizations, we obtain
\be
m_{p}^2 \simeq \left\{ \begin{array}{ll}
4\pi\alpha \left( n_e \langle 
{p_e^{-1}} \rangle + n_\gamma \langle 
{p_\gamma^{-1}}\rangle \right), & \mbox{QED} \\
{16\pi\over 3}\alpha_s \left( {n_q\over 3n_f} \langle 
{p_q^{-1}} \rangle + {n_g\over 8} \langle 
{p_g^{-1}}\rangle \right), & \mbox{QCD.} \end{array}\right.
\label{eq:plmass}
\ee

Near the BH horizon where the particle densities are high, the plasma
mass dominates over the vacuum mass, and leads to a significant
reduction in the \br cross section, since $\bar\sigma \propto
m_{th}^{-2}$.  Failure to take this into account would give a
misleading estimate of the onset of the photosphere in the case of
QED.

Several aspects of the implementation of thermal mass in the test
particle model should be mentioned. First, in eq.~(\ref{eq:plmass})
$n_e$ and $n_\gamma$ are the actual electron and photon densities
calculated analytically, whereas $p_e$ and $p_\gamma$ are kinetic
energies of the test particles, assumed to be relativistic. Secondly,
we recalculate the fermion thermal mass [using eq.~(\ref{eq:thmass})]
at each step in radius. Because $m_{th}$ changes, if the fermion
momentum was held fixed, its total energy $E=\sqrt{m_{th}^2+p_e^2}$,
would change.  Hence, we assume that the thermal mass correction is done
at the expense of momentum in such a way as to keep its total energy $E$
constant.

(The reader may wonder whether energy should indeed be conserved, due
to gravitational redshift.  In fact all significant redshifting of the
outgoing particle energies occurs within the first few Schwarzschild
radii of the horizon, where the classical concept of particles is not
yet well defined, due in part to the deBroglie wavelength being of the
same order as the Schwarzschild radius.  Moreover, the Hawking spectrum
(\ref{eqn:K1}) refers to energies as observed far from the BH. If we
were going to try to include redshift effects, we should blueshift the
initial particle energies accordingly, so that asymptotically they have
the usual distribution. This would only affect our simulation very
close to the horizon, which does not appear to be an important region
as regards the observable features of the photosphere.  Hence it is
simpler and seemingly a good approximation to ignore redshift.)

As in ref.~\cite{Heck:1}, we can estimate the thermal mass of an electron
in the QED photosphere using $n(r)=T_{BH}(3/2)^{\cN(r)}/(\pi^2 (4\pi)^2
r^2)$ for the densities of both photons and electrons, and
$3T_{BH}/(3/2)^{\cN(r)}$ for average particle energy at radius $r$.  Here,
$\cN(r)$ is the number of \br events an average particle undergoes between
the horizon and radius $r$. Since in each such scattering one new particle
is produced, $(3/2)^{\cN(r)}$ is the factor by which the total number of
particles has increased. In the test-particle model, since the total
number of particles $N(r)$ is known at each step, this increase is
calculated directly as $N(r)/N(r_h)$.  Then,
\be 
m_{th} \simeq
\sqrt{m_{e}^2+\frac{\alpha}{\pi(4\pi)^2}
    \frac{1}{r^2} \left(\frac{N(r)}{N(r_h)} \right) ^2}.  
\label{mthapprox}
\ee

\protect
\begin{figure}[h!]
\includegraphics[width=0.5\textwidth]{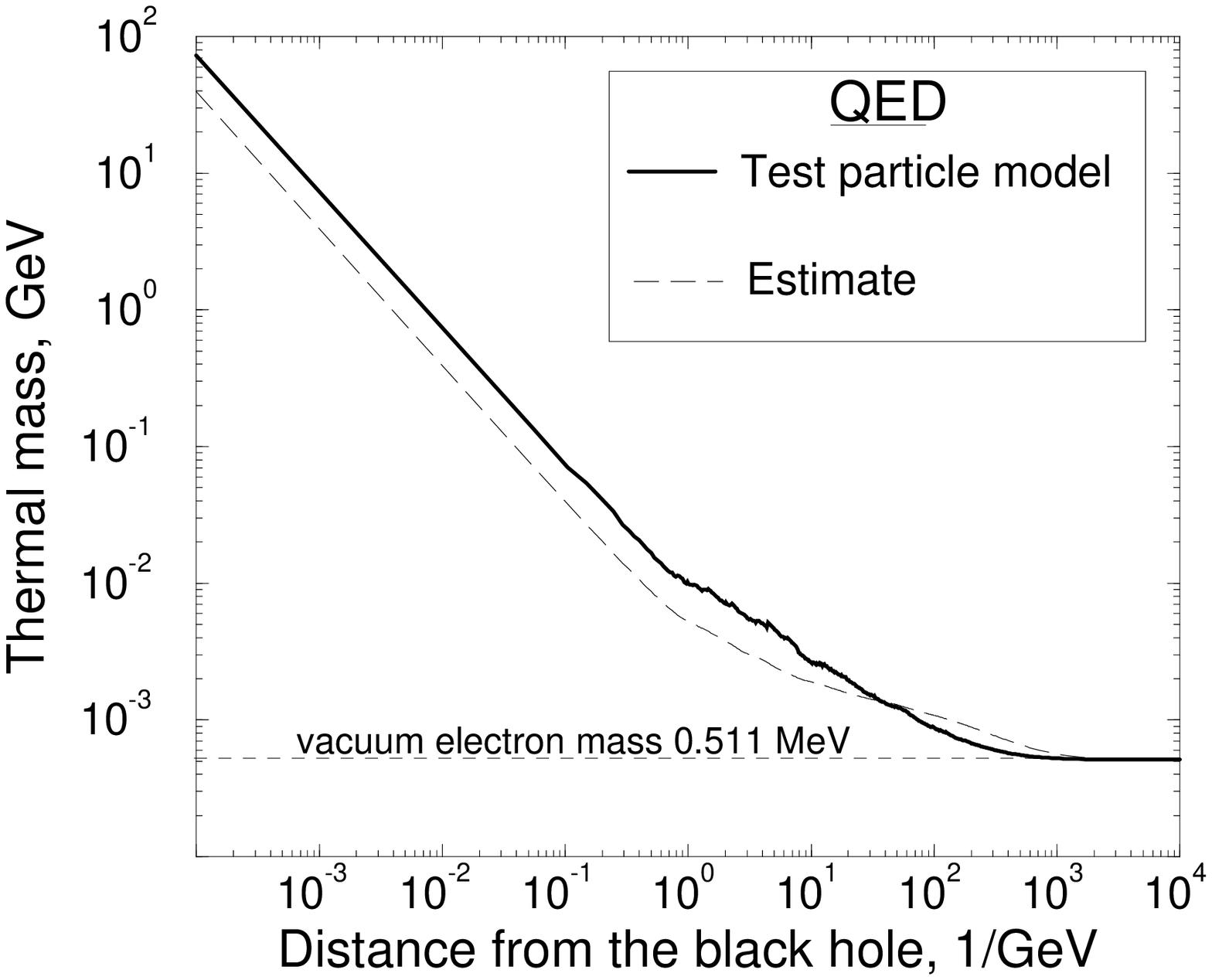} 
\includegraphics[width=0.5\textwidth]{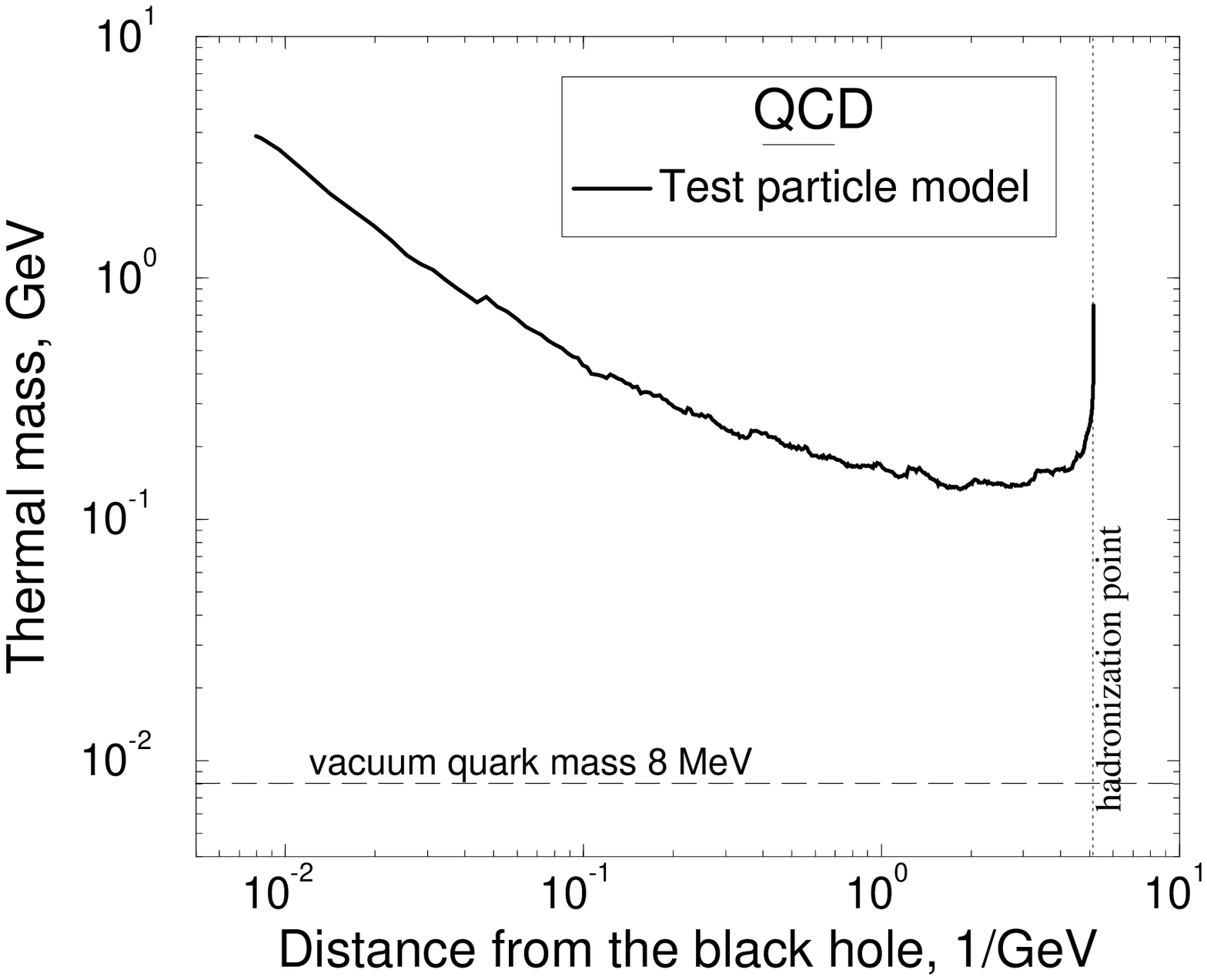} 
\caption{Thermal mass (and the estimate
(\ref{mthapprox})) in typical  (a) QED and (b) QCD photospheres.}
\label{fig:mass}
\end{figure}
\vspace{-1mm}

Fig.~\ref{fig:mass} shows how the thermal mass depends on radial distance
from the BH for two illustrative cases: the QED photosphere of a BH with
$T= 1000$ GeV, and the QCD photosphere for a $T= 10$ GeV black hole. We
see that in the QED plasma the factor of $1/r^2$ is dominant over any
growth in density as the radius increases, which causes the thermal mass
to decrease monotonically.  On the other hand, in the QCD plasma, the
particle production factor $N(r)/N(r_h)$ is more influential (due to
the running of the coupling constant with energy, in
particular) and just before hadronization starts, at average particle
energies $\langle E \rangle$ near the QCD scale $\Lambda_{QCD}$, the
thermal mass may again increase. Our perturbative formulas should not be
trusted quantitatively in this regime, however.  Fig.~\ref{fig:mass}a also
shows that the analytic approximation (\ref{mthapprox})  $m_{th}$ and the
numerically computed values are in reasonably good agreement.

\subsection{Interaction Distance and Relative Velocity}

The distance at which particles can interact via bremsstrahlung and pair
production is an important parameter, since we have to decide which test
particles may interact with one another and which may not.  In the heavy
ion collision problem, those test particles were allowed to interact which
were within a critical distance $b_c=\sqrt{\sigma/\pi}$ from each other. 
Notice that in that case, the relative velocity of the two particles was
typically large because the interacting nucleons were not moving parallel
to each other.  But in the present case of BH radiation, two particles
which are nearest neighbors typically are moving in almost the same
direction, {\it i.e.,} radially, leading to a suppressed relative velocity
$v_{rel}$.  Such particles have a small probability of interacting, since
they make a small contribution to the inverse mean free path,
eq.~(\ref{eqn:K9}).  The dominant interactions involve pairs of particles
with large relative velocity.  Near the horizon of the BH, such pairs
would consist of particles on opposite sides of the BH, moving away from
each other. 

Although classically it is somewhat counterintuitive to have
interactions between particles that have already passed each other, as
it were, there is no reason to exclude them so long as the particles
are still within the range of the interaction.  For \br it has been
shown \cite{RevModernPhysics} that at relativistic energies, small
momentum transfers of the order $k\simeq m_e$ contribute the bulk of
the total cross section.  Therefore the distance at which particles can
interact via bremsstrahlung (and pair production) is of order $m_e^{-1}$
in the vacuum.  In the plasma, accounting for the LPM effect, the
cutoff instead becomes the thermal mass $m_{th}^{-1}$.

Therefore in the radial interval $r_h < r < m_{th}^{-1}$ (assuming the
BH is microscopic so that the photosphere can indeed form), a given
particle in the plasma is always able to interact with some other
particle, with a large relative velocity, $v_{rel}\lsim 2$.  It is not
necessary to assume particles interact only with their nearest
neighbors, or to keep track of the exact spatial trajectories of the
particles, which simplifies our task of evolving the ensemble of test
particles.  

\begin{figure}[h!]
\centering
\includegraphics[width=0.5\textwidth]{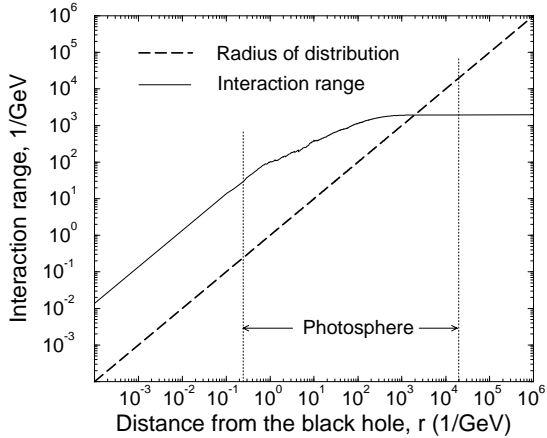} 
\caption{Maximum distance of interaction, $m_{th}^{-1}$, in a typical 
QED photosphere ($T_{BH}=1000$ GeV), indicating the rapid dissipation
of the photosphere at distances $r>m_{th}^{-1}$.} 
\label{fig:distance}
\end{figure}

Of course the above procedure no longer works at radii larger
than $m_{th}^{-1}$.  At that point we adopt the procedure of randomly
pairing particles, assuming the pairs represent nearest neighbors
separated by the average interparticle distance ${n(r)}^{-1/3}$,
and computing their actual relative velocity, given by the formula
\begin{eqnarray}
\label{eqn:K11}
v_{rel}&=&\frac{\sqrt{(p_1\cdot p_2)^2-m_1^2m_2^2}}{E_1E_2}\nonumber\\
 &\cong&
\sqrt{2\left(1-\frac{\mathbf{p_1} \cdot \mathbf{p_2}}{p_1p_2}\right)},
\end{eqnarray}
whose second form is valid in the relativistic limit.\footnote{This
definition of $v_{rel}$ arises from comparing the detailed form of the
collision term in the Boltzmann equation with that of the scattering
cross section. It is important to notice that the scattering angle
between $\mathbf{p_1}$ and $\mathbf{p_2}$ is evaluated in the rest
frame of the BH. This is the same frame in which the Boltzmann equation
is most naturally formulated for the present problem.} In this
large-radius regime we rely upon the randomization of velocities
provided by the scatterings themselves to keep the momenta from
becoming purely radial, which otherwise causes $v_{rel}$ to tend toward
zero.  Nevertheless the inevitable radialization of momenta as $r$
increases quickly overcomes the randomization due to scatterings,
causing the particles to become freely propagating and marking the end
of the photosphere.  This behavior is shown in Fig.~\ref{fig:distance}.
Although our matching between the $r<m_{th}^{-1}$ and $r>m_{th}^{-1}$
regions is somewhat crude, we believe it captures the essential physics,
and that a more accurate treatment, using detailed information about
each particle's trajectory, would only change the estimate of the 
photosphere's outer radius by a factor of order unity.
\protect

\section{Numerical Results}
\subsection*{QED photosphere}
This section will summarize the results of test particle 
simulation of black hole emission when only electrons, positrons 
and photons are taken into consideration.
This restriction is appropriate for black holes with $T_{BH} < 
\Lambda_{QCD}$. We will also treat higher temperatures in this context 
for the sake of understanding, deferring study of the effects of quarks 
and gluons until the next subsection. 

\subsection{Photosphere Formation}

We confirm the formation of a photosphere for black hole temperatures
$T_{BH} \sim 100$ GeV and higher, as was originally suggested in
\cite{Heck:1}. The photosphere first appears at a radius $r_0 \sim 10^4
\, r_h$ (where $r_h=1/4 \pi T_{BH}$ is the Schwarzschild radius), and
is characterized by a region of intense collisions terminating at a
distance of $r_f \sim 10^7-10^8 \, r_h$ from the black hole.  The
effectiveness of the collisions is demonstrated  by a very slow increase,
or even decrease, of the mean free path in the photosphere compared to
the interior and exterior regions.

\begin{figure}[h!]
\centering
\includegraphics[width=0.5\textwidth]{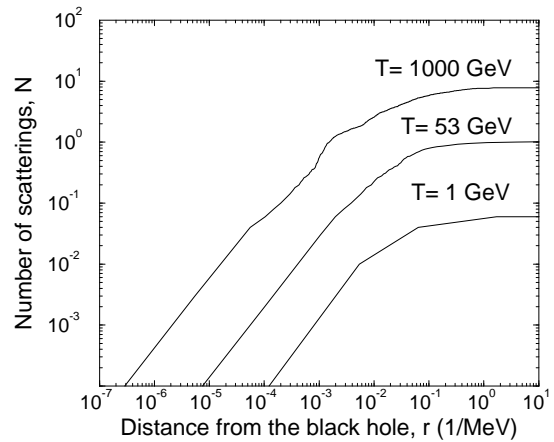}
\caption{The number of scatterings an average particle undergoes 
between the horizon and radius $r$: $\cN (r)$.}
\label{fig:cN} 
\end{figure}

The photosphere forms only for black holes above a certain critical 
temperature $T_c$. We use the same definition of $T_c$ as was 
introduced by Heckler \cite{Heck:1}. The idea is to define a quantity 
$\cN(r)$ denoting the number of scatterings an average particle undergoes between 
the horizon and some radius $r>r_h$.
The criterion for having a photosphere is taken to be that on average 
every particle undergoes a collision at least once between leaving the 
horizon and escaping to infinity, in other words that
\be
\lim_{r \to \infty} \cN(r) \geq 1.
\ee 
$T_c$ is then the temperature of a black 
hole for which this limit is exactly unity. The critical temperature we 
obtain is $T_c \simeq 50 \mbox{ GeV}$, whereas the result following from
the approximate method of ref.~\cite{Heck:1} is $ 45.2$ GeV. Our
determination of $\cN(r)$ 
is shown for several black hole temperatures in Fig.~\ref{fig:cN}.
\protect

\subsection{Photosphere Parameters}

Here we will present the photosphere parameters obtained from simulations
for approximately 30 different black hole temperatures.  These parameters
include the radii of the inner and outer surfaces, $r_i$ and $r_f$
respectively, the total particle production factor $P$ and the average
energy of the particles emerging from the photosphere $\Ef$. The latter is
relevant because it is the average energy of particles that may eventually
reach a distant observer. We will point out significant discrepancies
between these results and the fluid model used in \cite{Heck:1}, and
derive empirical formulas from our simulations showing the
dependence of $r_i$, $r_f$ and $\Ef$ on the BH temperature, $T_{BH}$.

\subsubsection*{Inner radius}

%
%
%
%
%
The radius of the inner surface of the photosphere ($r_i$) is defined 
by $\cN (r_i) = 1$, {\it i.e.,} the radius by which on 
\protect
\begin{figure}[!h]
\centering
\includegraphics[width=0.5\textwidth]{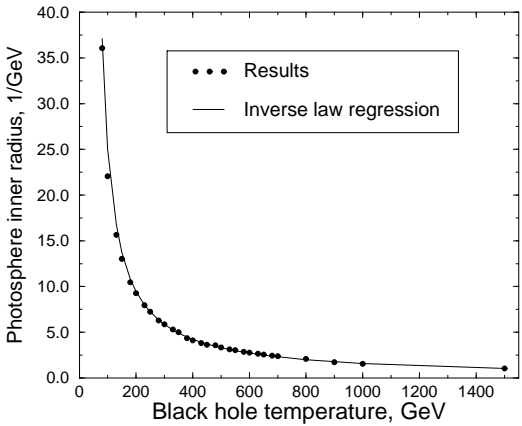}
\caption{Radii of inner QED photosphere surface for different black hole 
temperatures (black dots). Solid line represents inverse regression 
over these values, see Eq.~(\ref{eq:r0}).}
\label{fig:r0} 
\end{figure}
average every particle originating at the horizon has undergone one 
collision. 
The values of $r_i$ in units of 1 GeV$^{-1}\simeq 0.197$ fm are 
plotted in Fig.~\ref{fig:r0} as a function of black hole temperature.
As one can see from the graph, $r_i$ decreases with the temperature and 
is closely fitted by
\be
\label{eq:r0}
r_i=\frac{1}{\kappa T_{BH}},\qquad 
\kappa=(6.446\pm0.003) \times 10^{-4}.
\ee
We know that the radius of the Schwarzschild horizon is also inversely 
proportional to the black hole temperature $r_h=1/4\pi T_{BH}$, so that 
\be
r_i=\frac{4\pi}{\kappa} r_h \simeq 2 \times 10^{4} \, r_h.
\ee
By this criterion, the photosphere starts to develop much closer to the 
black hole than was predicted ($r_i \sim 10^9 r_h$) in~\cite{Heck:1} 
using a fluid model for the interacting particles.

\subsubsection*{Edge Radius}

The outer radius of the photosphere, $r_f$, can be defined as the distance
after which particles effectively free stream away without significant
interactions with one another.  The mean free path $\lambda$ quickly
begins to diverge at this point.  
\protect 
\begin{figure}[h!] 
\centering
\includegraphics[width=0.5\textwidth]{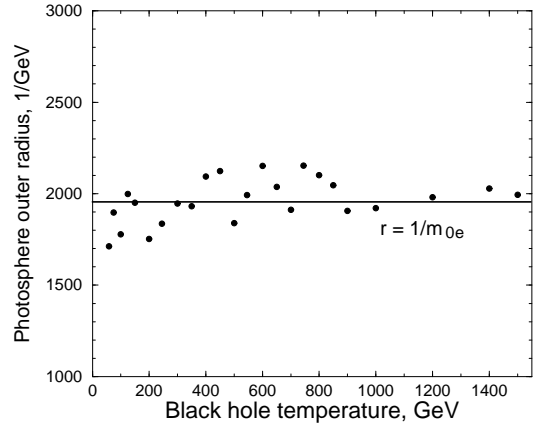} 
\caption{Radii of outer QED
photosphere surface for different black hole temperatures. The solid line
shows $r=1/m_{e}$, the radius beyond which particles can no longer
interact with each other by \br.} 
\label{fig:rf} 
\end{figure} 
If $\lambda_i$ is the mean free path at step $i$ then the condition
$\lambda_{i+1}/\lambda_{i} \gg \lambda_i/\lambda_{i-1}$ is a convenient
criterion to define the end the photosphere, and $r_f$ is just the radius
the particles have reached by step $i$.  This value is easily found in
practice because in the next step after $r_f$, $\lambda$ is usually
several orders of magnitude larger than its previous value, implying that
the particles have become virtually free.  Because $\lambda(r)$ is
diverging so quickly near $r_f$, the latter is inherently difficult to
determine precisely, even when the steps size is made smaller near the
edge of the photo\-sphere. Great precision in $r_f$ is not essential,
however, because it is not an observable quantity. 

The values of $r_f$ remain in the same range (1700--2200 GeV$^{-1}$) for
all black hole temperatures shown and are consistent with being
independent of $T_{BH}$ at the horizon.  Only the statistical
dispersion around the mean value decreases with rising temperature
(Fig.~\ref{fig:rf}).  The mean value is approximately $r_f \sim 2000$
GeV$^{-1}$, which is close to $1/m_{e}$.  This is easy to understand in
light of our previous discussion (section 3.3 and fig.~3)  of the range of
the \br interaction, whose maximum value is of order $1/m_e$.  Because the
trajectories of particles within the range of interaction rapidly become
parallel at radii beyond this cutoff, the photosphere quickly dissipates.

\subsubsection*{Particle Production}

\begin{figure}[h!]
\centering
\includegraphics[width=0.5\textwidth]{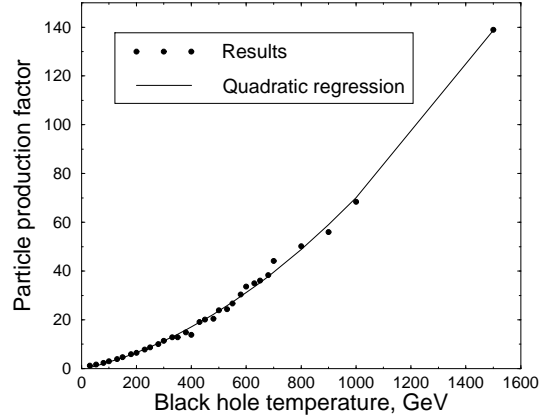}
\includegraphics[width=0.5\textwidth]{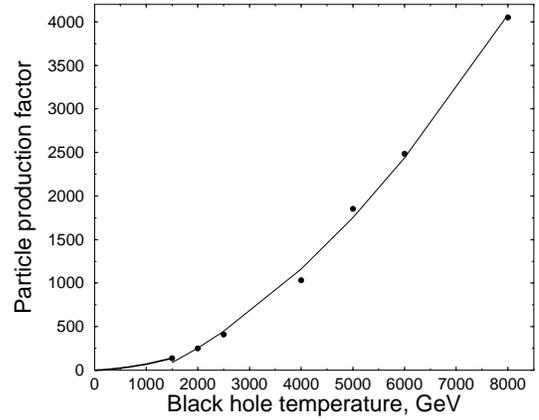}
\caption{Particle production factor $N(r_f)/N(r_h)$ versus $T_{BH}$ for
(a) $T_{BH} < 1.5$ TeV and (b) $T_{BH} < 8$ TeV. 
Solid line is the quadratic fit to the results [eq.~(\ref{eq:pp})].}
\label{fig:pp} 
\end{figure}

Another useful parameter for characterizing the photosphere is the
total particle production factor, given by $P=N(r_f)/N(r_h)$,
where $N(r)$ is the number of test particles at radius r. $P$ can be used 
to quantify the probability of interactions inside the photosphere, since 
at each collision $N\rightarrow N+1$. Figure~\ref{fig:pp} shows $P$ as a 
function of black hole temperature.
\protect
We find that it can be represented by a quadratic fit (solid line):
\be
P(T_{BH}) \simeq a T_{BH}+ b T_{BH}^2,
\label{eq:pp}
\ee
where 
\begin{eqnarray}
a = \left\{ \begin{array}{ll}
         0.026 \mbox{ GeV}^{-1}, & T_{BH} < 2 {\rm\ TeV};\\
                        \\
         0.226 \mbox{ GeV}^{-1}, & T_{BH} > 2 {\rm\ TeV};\\
             \end{array}
     \right.
\nonumber
\\
\nonumber
\\
b = \left\{ \begin{array}{ll}
         4.5\times 10^{-5} \mbox{ GeV}^{-2}, & T_{BH} < 2 {\rm\ TeV};\\
                        \\
         4.2\times 10^{-5} \mbox{ GeV}^{-2}, & T_{BH} > 2 {\rm\ TeV}.\\
             \end{array}
     \right.
\end{eqnarray}
The dominant quadratic term implies that the particle number density 
in the photosphere increases rapidly in the last stages of black hole
evaporation, as $T$ approaches the Planck scale.
The slightly different dependences for $T_{BH}>2$ TeV and $T_{BH} < 2$
TeV can be understood by the argument of section 3.1: this is the
temperature where randomizing effects of elastic scattering become
relevant.  However, the effect is not dramatic.

\subsubsection*{Average Final Energy}

\begin{figure}[h!]
\centering
\includegraphics[width=0.5\textwidth]{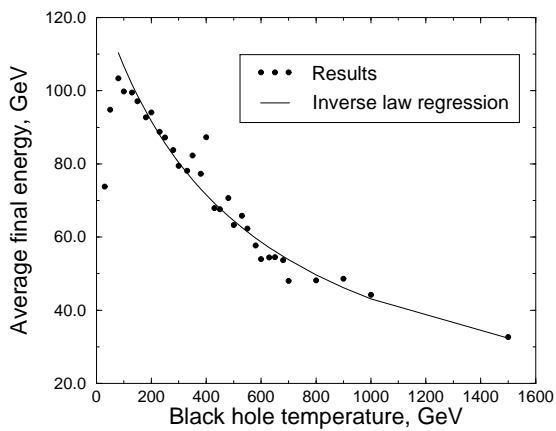}
\includegraphics[width=0.5\textwidth]{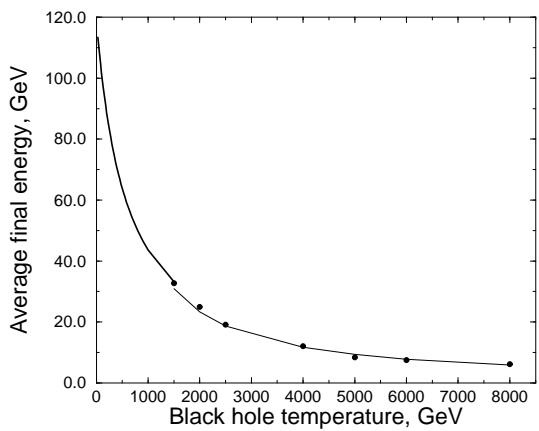}
\caption{$\Ef$, average particle energy at the outer edge of the QED
photosphere, for (a) $T_{BH} < 1.5$ TeV and (b) $T_{BH} < 8$ TeV.}
\label{fig:Ef} 
\end{figure}

From an observational point of view, the reduction in average energy of
emitted particles is one of the most relevant consequences of the
photosphere. At the horizon, $\overline{E}_i$ is approximately $3.1
T_{BH}$.  But the photosphere can reduce this number dramatically, so
that a distant observer sees a much softer spectrum.  Our results for
the $T_{BH}$-dependence of the average energy at the edge of the
photosphere, $\Ef$, are displayed in Figure~\ref{fig:Ef}.  These values
are significantly higher than those ($\sim 1$ MeV) found in
ref.~\cite{Heck:1}, and the temperature-dependence is also very
different. This apparently stems from the use of a fluid description
in \cite{Heck:1} which is not really applicable.

\protect

We can compute $\Ef$ from $\overline{E}_i$ and the particle production 
factor, Eq.~(\ref{eq:pp}): 
\begin{table*}[!t]
\centering
\begin{tabular}{|r|c|c|c|}
\hline
 & & & \\
$T_{BH}$ & $r_i$ & $r_f$ & $\Ef$ \\
\hline
 & & & \\
60 GeV   & $1.5 \times 10^3 \mbox{\ fm}$ 
& $8.6\times 10^3\mbox{\ fm}$
                             & $97.0$ GeV \\
 & & & \\
300 GeV  & $30\mbox{\ fm}$ & $9.6\times 10^3\mbox{\ fm}$
                             & $79.5$ GeV \\
 & & & \\
1000 GeV & $7.9 \mbox{\ fm}$ & $9.6 \times 10^3 \mbox{\ fm}$ 
                             & $44.2$ GeV \\
 & & & \\
\hline
\end{tabular}
\caption{QED photosphere parameters (temperature at horizon, inner
and outer photosphere radii, and average particle energy at the outer
edge of the photosphere)
for several black hole temperatures 
obtained in test particle simulations.} 
\label{tb:me}
\end{table*}
\begin{eqnarray}
\Ef&=&\frac{\overline{E}_i}{P} = \frac{(2.7 {n_b\over n} + 3.15 
{n_f\over n})T_{BH}}
{a T_{BH}+ b T_{BH}^2}\nonumber\\
&=&\frac{3.1}{b+cT_{BH}},
\end{eqnarray}
where $n_b$, $n_f$ and $n$ are the respective densities of bosons,
fermions, and all particles at the horizon.
In the limit of high black hole temperatures \mbox{($a\ll b T_{BH} 
\Leftrightarrow 
T_{BH} > 10^4$ GeV)}:
\be
\Ef \simeq 7.4\times10^4 \frac{\mbox{GeV}^2}{T_{BH}}
\label{eq:Ef}
\ee
This is a remarkable result since it predicts that for black holes of
temperature $T_{BH} \gsim 100$ GeV ($M<10^{12}$g) the effective
temperature of emitted particles (that which would be inferred by
observers far away from the hole) is actually lower, the higher is their
temperature at the horizon.  For an individual black hole, which is losing
mass and hence becomes hotter, this means that its apparent temperature
goes down.  The word ``temperature'' is used loosely here: the spectrum
is nonthermal, with a higher luminosity at low frequencies than that of
a blackbody (see fig.~10 below).

Eq.~(\ref{eq:Ef}) was derived on the assumption that the black hole
horizon temperature does not change much on the time scale of particle
diffusion from the horizon to the outer edge of the photosphere.
Nevertheless, we expect the qualitative picture to be the same even for
black holes temperatures above this limit of validity.  It should also
be kept in mind that we are discussing only electromagnetically
interacting particles so far.  The behavior of the QCD photosphere
(below) is quite different.

The time development of an individual BH, based on the above results, can
now be described.  As far as only QED emission is concerned, a photosphere
starts to develop around the black hole when it evaporates to a mass of
$M~\simeq~5\times~10^{12}$g. At this point the average energy of emitted
particles, instead of going up inverse proportionally to the BH mass,
levels off and begins to decrease. On the other hand, the total luminosity
increases, and the spectrum becomes softer than that of a black body. The
outer edge of the photosphere remains at a radius of 400 fm $\sim
m_e^{-1}$.  However, its inner radius, $r_i \sim 700 \, r_h$, shrinks with
the Schwarzschild horizon $r_h$.  Eventually, if the steady-state
assumption remains valid at these temperatures, the edge will cool to
$\Ef\sim m_{e}$, when the positrons and electrons annihilate and
no further cooling occurs. At this point, however, the black hole is
within $10^{-10}$ s of its total evaporation. 

To give some idea of how the results of the test particle method differ
from the estimates in \cite{Heck:1}, where a nonperfect fluid model
was used, we tabulate the relevant quantities in table~\ref{tb:me}, for
three different BH temperatures. 

\subsection{Inside the Photosphere}

\protect
\begin{figure}
\includegraphics[width=0.4\textwidth]{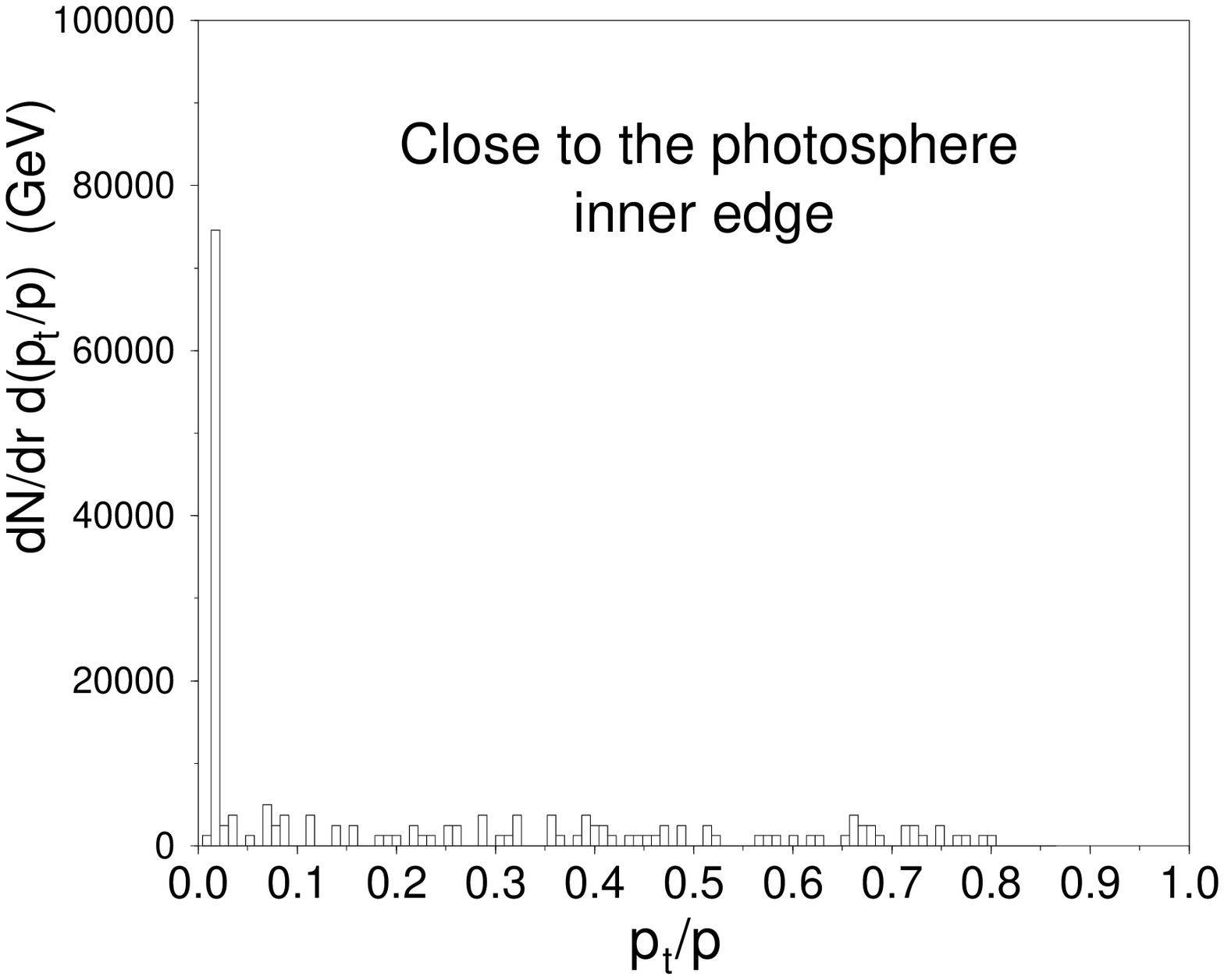}
\includegraphics[width=0.4\textwidth]{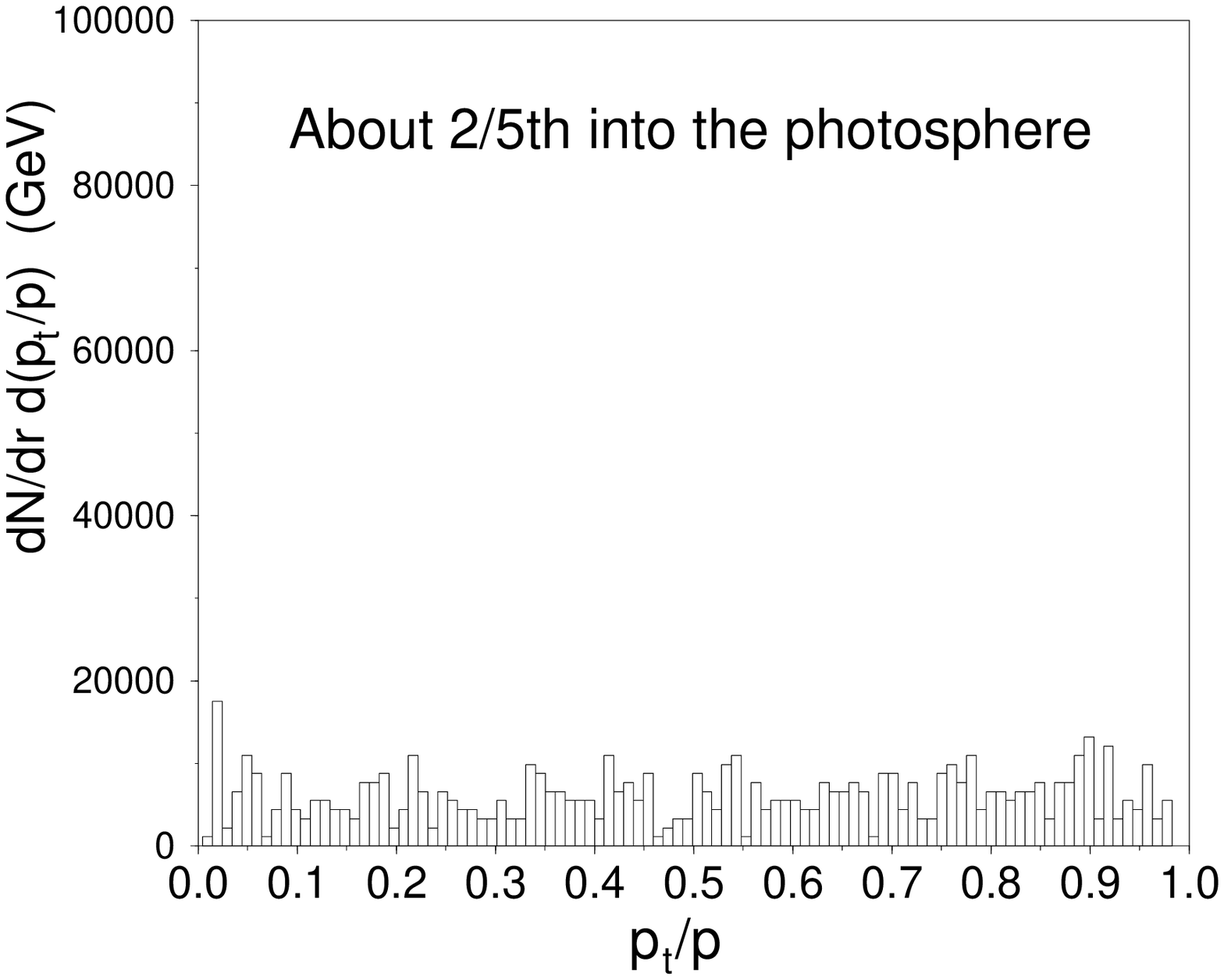} 
\includegraphics[width=0.4\textwidth]{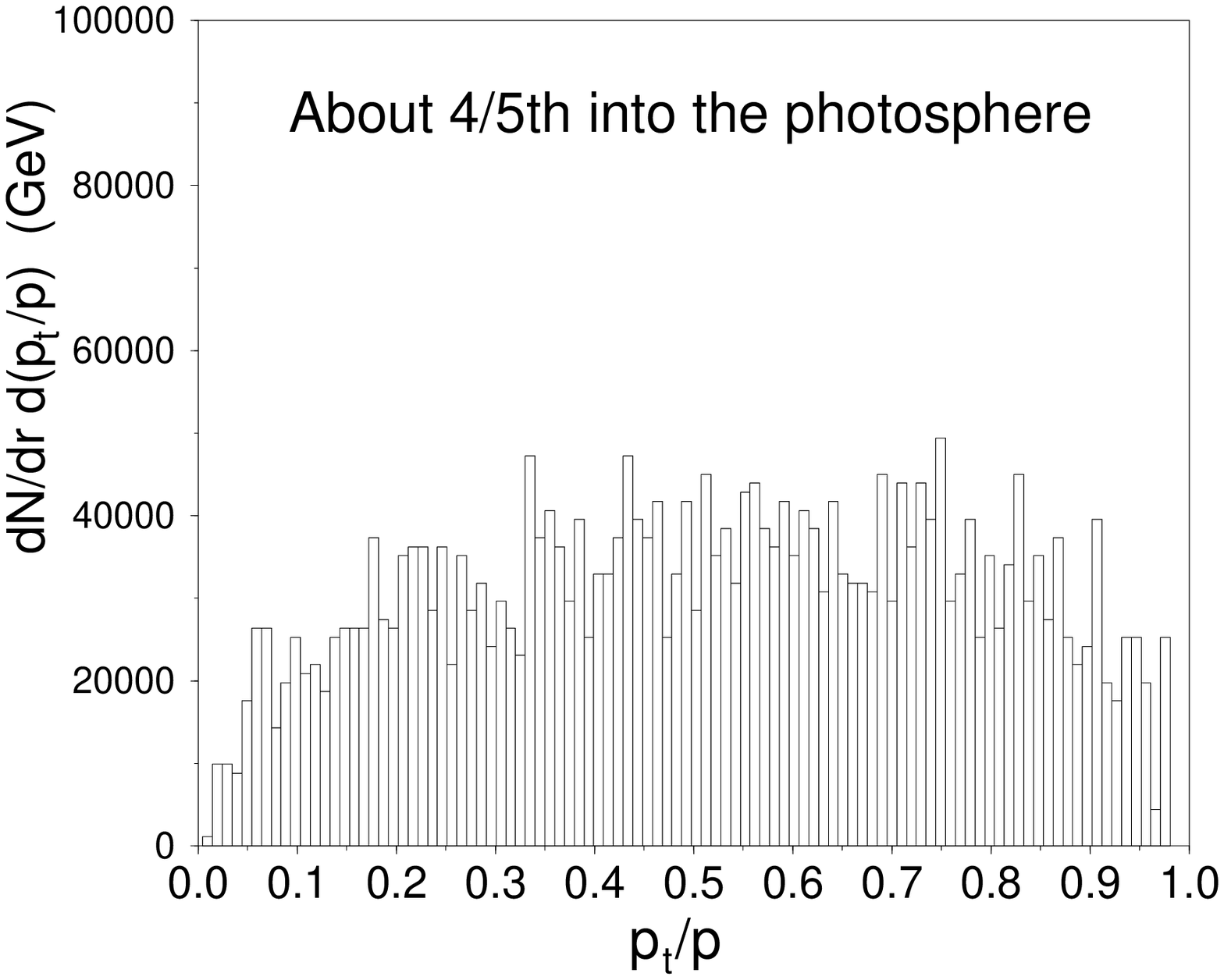}
\includegraphics[width=0.4\textwidth]{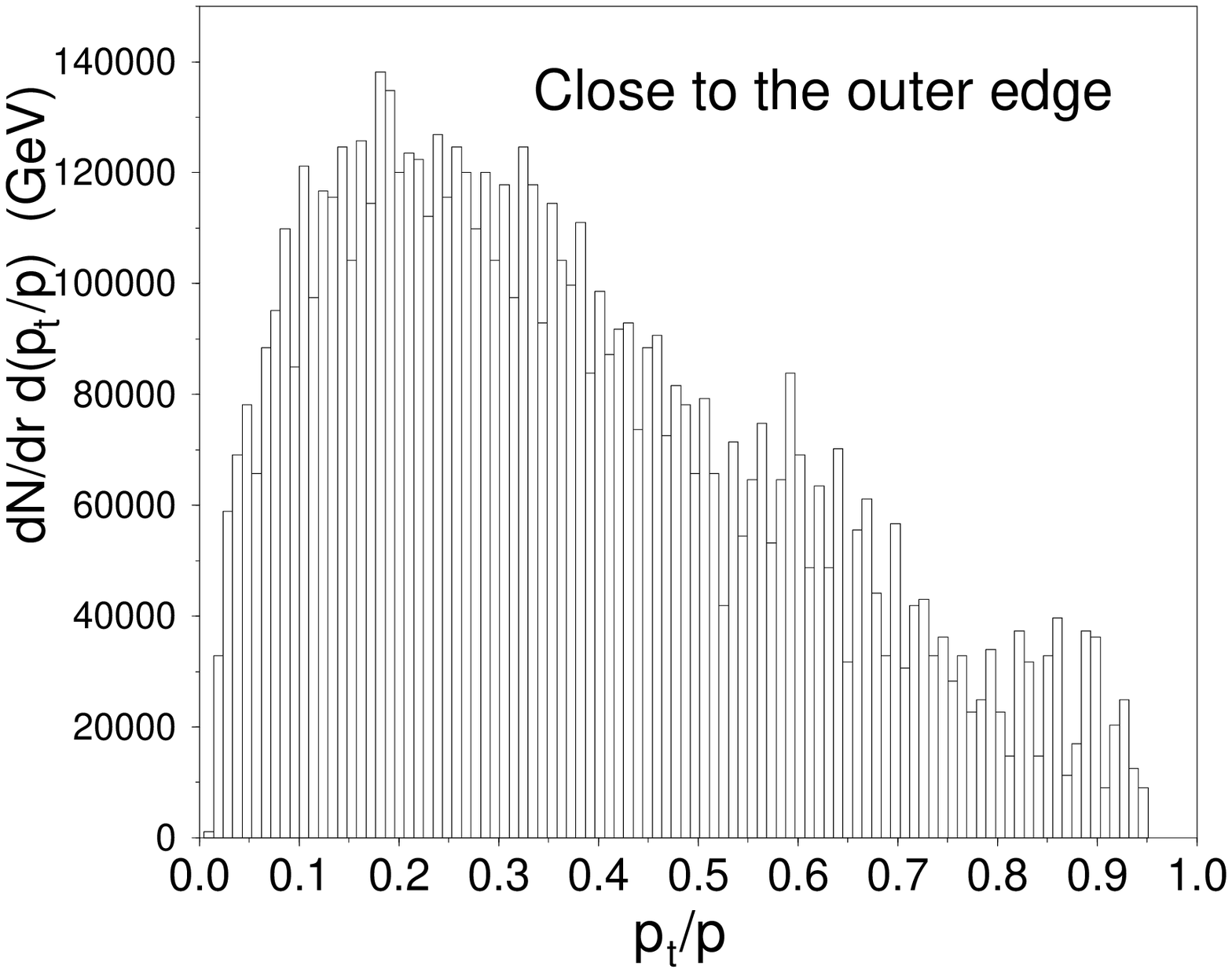}
\caption{Distribution of $p_t/p$ for several radial shells
 in the QED photosphere of a BH with $T=1000$ GeV.  }
\label{fig:pt}
\end{figure}
%
A more detailed picture of the plasma can be seen from the
particle momenta distributions at different radii inside the photosphere. 
Since no significant interactions start before $r \sim 10^4 \, r_h$, 
particles move increasingly radially until the inner boundary of the 
photosphere.  By this point they have on average undergone one 
interaction and their momenta begin to get randomized.
Close to the end of the photosphere, the mean free path 
increases the radial components of the momenta again start to increase. 
These developments can be seen in the distributions of the transverse
momenta.  In fig.~\ref{fig:pt} we show the distributions of the fraction
of momentum which is transverse to the radial direction,
$p_t/p = p_t/\sqrt{p_t^2+p_r^2}$, for several radii 
inside a sample photosphere.
These figures show the overall growth in
particle density as well as the shift to larger $p_t$ as one goes from
the inner boundary of the photosphere toward its interior.

\subsection{Spectra}

\begin{figure}[b!]
\includegraphics[width=0.4\textwidth]{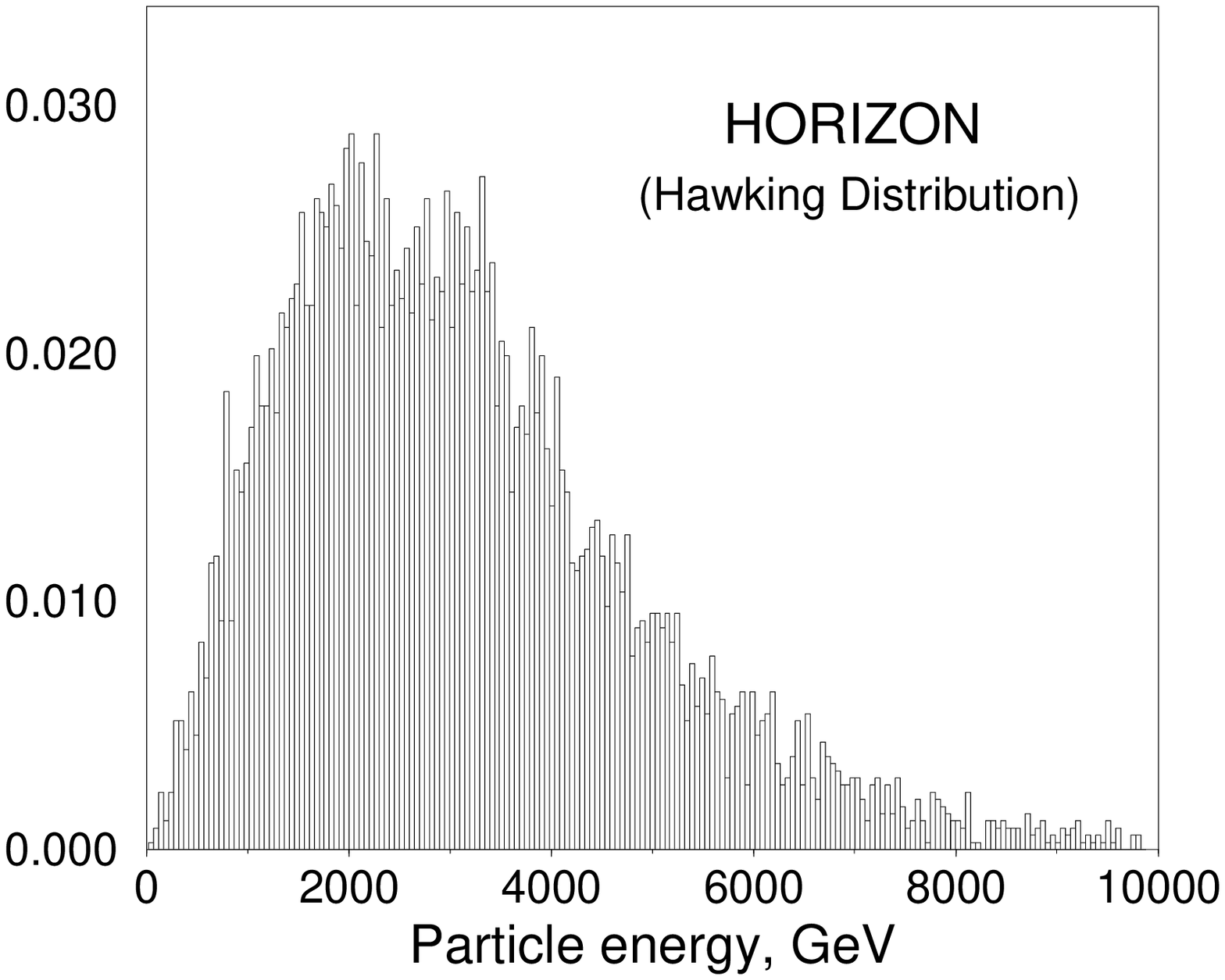}
\includegraphics[width=0.4\textwidth]{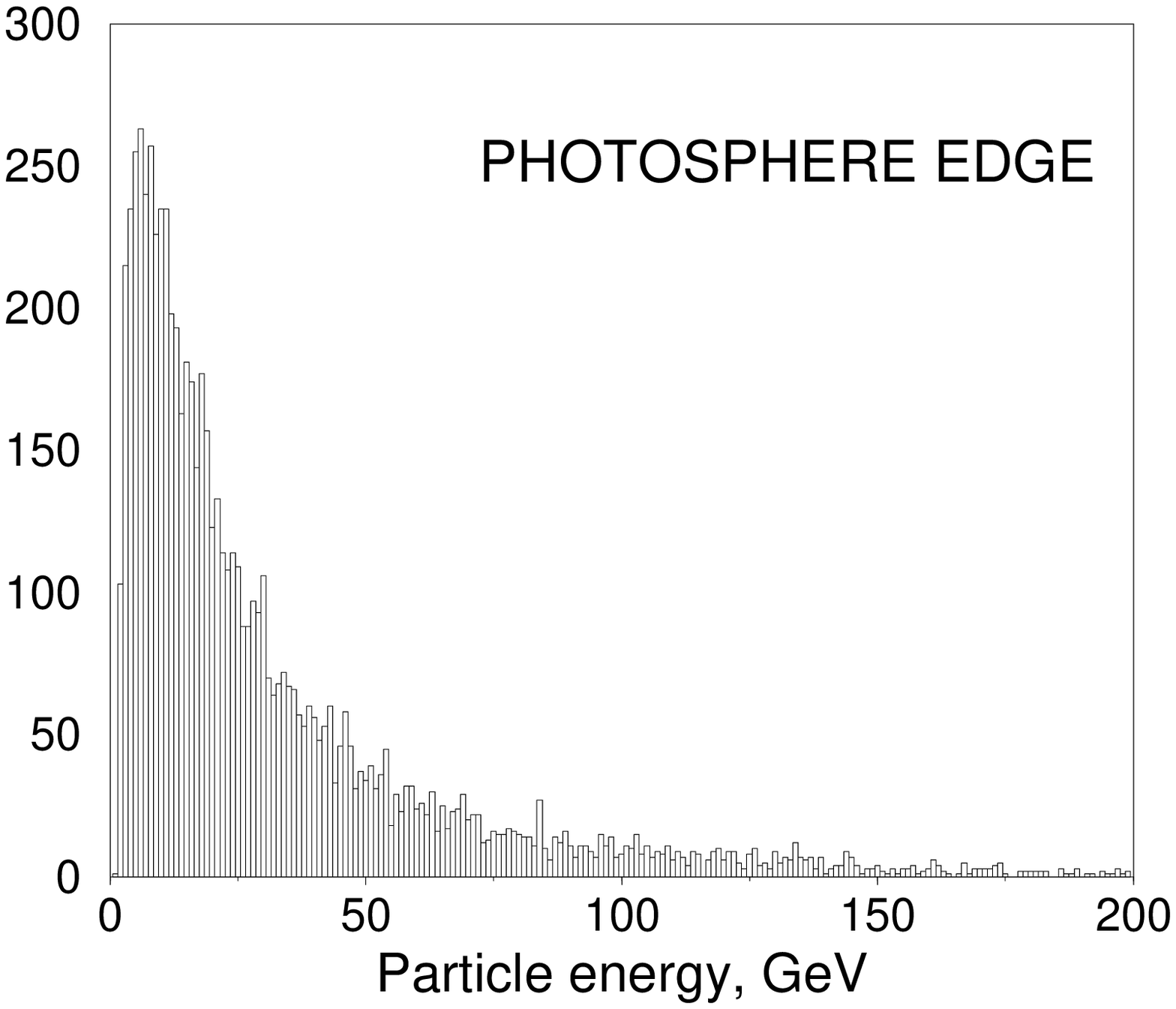}
\caption{Energy spectrum of a 1000 GeV ($10^{11}$g) black hole at the 
Schwarzschild horizon and near the edge of the QED photosphere.} 
\label{fig:spectra} 
\end{figure}

\begin{figure}[t!]
\centering
\includegraphics[width=0.5\textwidth]{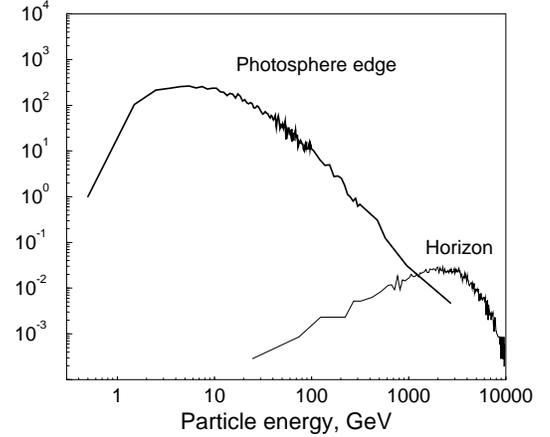}
\caption{Same as figure \ref{fig:spectra}, but with the two spectra
 superposed.} 
\label{fig:logspectra} 
\end{figure}

Finally, we examine the particle energy spectra at the 
black hole horizon and at the photosphere edge (Fig.~\ref{fig:spectra}).
The shift toward lower energies is the most significant difference
between the original distribution at the horizon and those within the 
photosphere. The two are shown together in Figure \ref{fig:logspectra}, 
where it can be seen that the softening in energy is accompanied by an 
increase in the number of particles, as is required by energy conservation.
\protect
%
%

The QED photosphere by itself serves as a kind of toy model for realistic
black holes, which are also emitting quarks and gluons at the temperatures
we are considering. We now turn our attention to the QCD photosphere. 

\subsection*{QCD photosphere}
%

A black hole whose temperature exceeds $\Lambda_{QCD} \sim 200$ MeV,
thus having a mass $M_{BH} \lsim 5 \times 10^{14}$ g, emits quarks and
gluons which, as proposed by MacGibbon and Weber in ref.~\cite{MacWeb},
fragment into hadrons, decaying in their turn into stable particles. We
have modeled the interactions of the quarks and gluons before
hadronization takes place. In this regime, as suggested by Heckler in
ref.~\cite{Heck:1}, a quark-gluon plasma similar to the electron-photon
photosphere in QED may develop, analogously changing the energy
spectrum of the particles.

\subsection{Test Particle Method in QCD}

Our treatment does not attempt to give a detailed model of QCD
interactions after hadronization begins. However, the onset of the
photosphere can be established in terms of free quarks and gluons
interacting with each other. We assume that hadronization occurs at a
distance of $\sim \Lambda_{QCD}^{-1}$ and that no significant softening
of the particle spectrum occurs after this point. Hence, we make the
same test particle simulation as in QED, only with different
interaction cross section, masses, and number densities. Due to the
much larger coupling constant and greater number of degrees of freedom,
we expect the photosphere to develop at temperatures $T_{BH}$ far below
the critical temperature for QED, $T_c \sim 50$ GeV, and to reach
higher densities than in the QED case.

\begin{figure}[h!]
\centering
\includegraphics[width=0.1\textwidth, height=0.045\textheight] 
{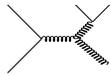} 
\caption{Nonabelian contribution to quark-gluon \br.}
\label{fig:irrelevant}
\end{figure}

To investigate the photosphere in QCD, we recall that the key 
ingredient was the inclusion of $2\rightarrow 
3$ body interactions in the collision term of the Boltzmann equation 
(\ref{BEq-r}). Like electrons and 
photons, quarks and gluons can also interact via bremsstrahlung ($qq 
\rightarrow q\,q\,g$) and pair production ($qg \rightarrow 
q\,q\,\overline{q}$). The dominant diagrams (which give large logarithms
in the cross section at low momentum transfer due to $t$-channel
propagators) are the same as for QED,
fig.~\ref{fig:relevant}. 
In addition, there are intrinsically nonabelian contributions like  
fig.~\ref{fig:irrelevant}
\vspace{0.3cm}
\protect
\vspace{-0.2cm}
which we do not expect to be parametrically larger than those of
fig.~\ref{fig:relevant}. We will therefore make the simple
approximation of modeling the \br and pair production cross sections
using the same form as in QED, but replacing the fine structure
constant by $\alpha_s$ and the electron mass by the mass of the 
lightest quark:
\be
\sigma_{brem}^{QCD} \simeq \frac{8\alpha_s^3}{m_{th}^2}\ln \frac{2E}
{m_{th}}.
\ee
Here $m_{th} = \sqrt{m_q^2 + m_p^2}$ using the QCD plasma mass,
eq.~(\ref{eq:plmass}), and we take $m_q$ to be the average of the up
and down quark masses, $m_q \cong 8$ MeV.  Based on the previous QED
results, heavier quarks are expected to make a subdominant contribution
to the photosphere, since their \br cross sections are smaller by a 
factor of the mass ratio squared.  The dependence of the thermal quark
mass on radial distance in the photosphere was already shown in 
fig.~\ref{fig:mass}.

In contrast to QED, the QCD coupling constant depends 
strongly on energy. To leading order in perturbation 
theory~\cite{Halzen:Martin},
\be
\alpha_s(\mu)=\frac{12\pi}{\left(33-2n_f \right)\ln \left( 
\mu^2/\Lambda_{QCD}^2 \right)}, 
\ee
where in our radial evolution we take $\mu$ to be the average particle
energy at a given radius, $\Lambda_{QCD} \sim 200$ MeV, and $n_f$ ranges
between 3 and 5 for the BH temperatures we are considering, depending on
the number of quark species with masses less than the average energy at a
given distance from the horizon. 

When the effective temperature of the quark-gluon plasma becomes of
order $\Lambda_{QCD}$, perturbation theory in $\alpha_s$ breaks down,
and the quarks and gluons hadronize. This process will be discussed in
Section 4.7.  At distances greater than $\Lambda_{QCD}^{-1}$, instead
of a gas of quarks and gluons, we will have a plasma of pions and
nucleons, which can in principle continue to cool through pion \br ($n
n \to n n \pi$ or $\pi\pi\to \pi\pi\pi$).  However the residual strong
interactions of pions and nucleons are screened, relative to those of
the quarks and gluons, by confinement.  Also the relevant scale for
the range of pion \br is $m_\pi^{-1}$, which is much shorter than the
range of quark-gluon brehmsstrahlung.  We expect these two reductions
in the strength of $2\to 3$ scattering processes to make the hadrons
much less effective than quarks and gluons in perpetuating the
photosphere.  Thus one might anticipate that the QCD photosphere ends
rather abruptly beyond distances of order $\Lambda_{QCD}^{-1}$.

To semiquantitatively investigate this post-hadronization regime, we 
modeled the hadron gas by replacing $\alpha_s$ by the pion-nucleon
fine-structure constant, $f^2/4\pi \cong 0.09$, obtained from
low-energy pion-nucleon scattering \cite{Barnes}; we also substituted
the quark mass with the pion thermal mass, which we estimate
analogously to eqs.~(\ref{eq:thmass}) and (\ref{eq:plmass}).  The
result is that hadron-hadron interactions are indeed ineffectual for
prolonging the photosphere.  Henceforth we simply use the hadronization
criterion to determine where the photosphere ends.

%
%

\subsection{Parameters of QCD Photosphere}

\begin{figure}[h!]
\centering
\includegraphics[width=0.5\textwidth]{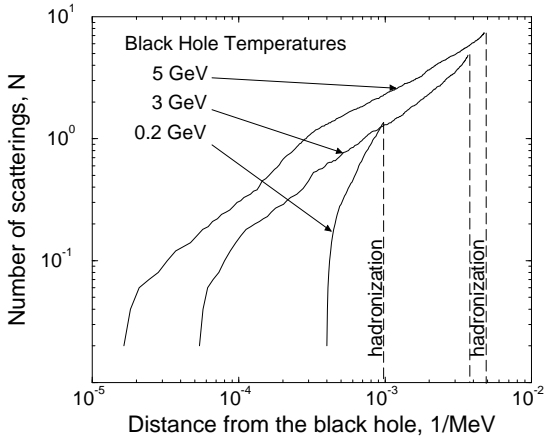}
\caption{Average number of scatterings $\cN$ in QCD photosphere as a 
function of radius $r$ for several black hole temperatures.} 
\label{fig:cNqcd}
\end{figure}

We have found that the QCD photosphere starts to develop for any black
hole whose temperature exceeds a critical value
\be
T_{c} \simeq 175 \mbox{ MeV.}  
\label{eq:TcrQCD}
\ee
This is more than two orders of magnitude lower than the critical
temperature for the QED photosphere. It agrees with the analytical
estimate in \cite{Heck:1}, $T_{c} \sim \Lambda_{QCD}$.  Recall
that $T_c$ is defined to be the temperature at which each particle on
average undergoes one scattering during its outward propagation.  In
Fig.~\ref{fig:cNqcd} we show the average number of scatterings per
particle as a function of distance, $\cN(r)$, at several BH
temperatures.  In contrast to the QED case (fig.~\ref{fig:cN}), where
$\cN(r)$ levels off at a universal value of the
final radius $r_f$ marking the end of the photosphere, in QCD $r_f$ 
depends strongly on the temperature. The radius at which hadronization
takes place grows with $T_{BH}$, which will be quantified below.
\protect

\begin{figure}[h!]
\centering
\includegraphics[width=0.5\textwidth]{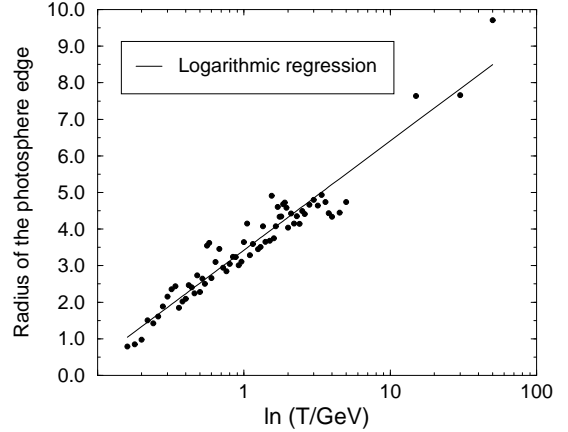}
\caption{Radius of the outer surface of the QCD photosphere ($r_f$) versus 
logarithm of the black hole temperature.}
\label{fig:rfqcd}
\end{figure}

The precise definition of the final radius, $r_f$, differs in the QCD
case from that which we used for QED: for QCD we must decide just where
hadronization occurs.  One can imagine several possible criteria:  when
the interparticle spacing ${[n(r_f)]}^{-1/3}$ begins to exceed
$\Lambda_{QCD}^{-1}$;  when the average particle energy
$\overline{E}(r_f)$ becomes of order $\Lambda_{QCD}$;  or when the
coupling constant $\alpha_s(r_f)$ becomes of order unity. We find that
all three are roughly equivalent in that the value of $r_f$ depends
only marginally on which one is used.  We adopt the coupling constant
criterion to define $r_f$ in the results that follow.
Fig.~\ref{fig:rfqcd} shows that $r_f$ is well described
 by a logarithmic growth in the photosphere radius with the black hole
temperature:
\be
r_f=A+B\ln\left(\frac{T_{BH}}{1\mbox{ GeV}}\right),
\label{eq:rfqcd}
\ee
where $A=3.25\pm 0.09\mbox{ GeV}^{-1} \simeq 0.65$ fm and $B=1.45\pm 
0.06 \mbox{ GeV}^{-1} \simeq 0.29$ fm.
\protect

QCD also differs from QED for the onset of the photosphere, whose
inner radius is denoted by $r_i$.  We find that significant QCD
interactions begin quite close to the horizon: $r_i\sim r_h$ for the
whole range of BH temperatures for which the QCD photosphere forms.

\begin{figure}[h!]
\centering
\includegraphics[width=0.5\textwidth]{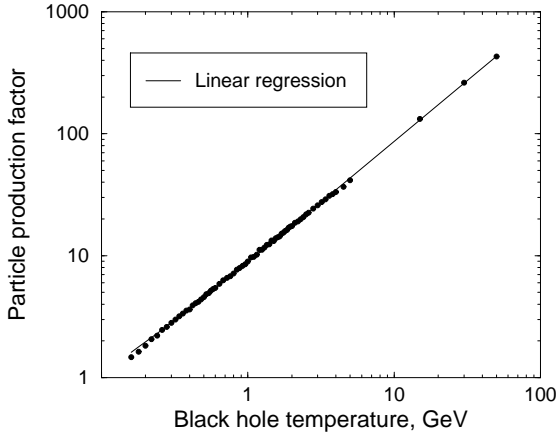}
\caption{Total particle production in the QCD photosphere versus black hole 
temperature.}
\label{fig:pqcd}
\end{figure}

The parameter which best characterizes the intensity of interactions in 
the QCD photosphere is the total particle production factor,
$P(T_{BH})=N(r_f)/N(r_h)$. As shown in Fig.~\ref{fig:pqcd}, we find that 
it increases quite linearly with black hole temperature:
\be
P(T_{BH}) = (8.62\pm0.01) \frac{T_{BH}}{\mbox{GeV}}.
\label{eq:ppQCD}
\ee
This, again, contrasts with the QED case, which displayed a noticeable
quadratic dependence on $T_{BH}$ at high temperatures.
\protect

The final spectra at the photosphere surface are not exactly thermal,
but they can be fitted over most of the range where $dN/dE$ is significant
by a Boltzmann distribution,
\be 
	{dN\over dE}\propto \exp(-E/T_0),
\label{qcd_fit}
\ee
where the effective temperature at the photosphere is
independent of $T_{BH}$:
\be
	T_0 = 300 {\rm\ MeV}.
\ee
\protect
\begin{figure}[h!]
\centering
\includegraphics[width=0.5\textwidth]{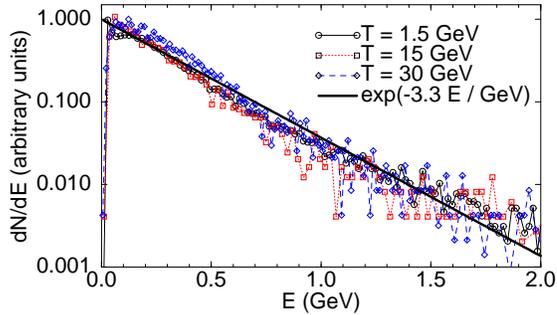}
\caption{Linear-log plot of QCD photosphere spectra (normalized to have
same peak value) for three different horizon temperatures, and the
analytic fit $dN/dE = \exp(-E/300$ MeV).}
\label{fig:qcd_dist}
\end{figure}
In fig.~\ref{fig:qcd_dist} we show the superimposed spectra at the
photosphere for three
different initial temperatures, $T_{BH} = 1.5$, 15 and 30 GeV, along with
the fit (\ref{qcd_fit}).  The spectra are normalized to have the
same maximum value so the shapes can be compared.  They rise very
sharply from $dN/dE = 0$ at $E=0$.  The actual normalization of the
flux $dN/dE dt$ will be discussed below.

From our simulations at different temperatures we can reconstruct the
time evolution of the QCD photosphere of an individual BH.  A black
hole that has reached a temperature greater than $T_c = 175$ MeV,
corresponding to a mass $M \lsim 5\times 10^{14}$ g, emits quarks
and gluons which almost immediately begin interacting to form a
photosphere very close to the horizon, $r_h \lsim 0.1$ fm. As the black
hole temperature continues to rise, the photosphere inner radius
shrinks along with the horizon ($r_i \sim r_h = 1/4\pi T_{BH}$), while
the outer radius grows logarithmically with $T_{BH}$.  Particles
emitted from the horizon with average energy $\overline{E}_i \sim 3\,
T_{BH}$ fragment in the photosphere into lower energy particles. The
higher $T_{BH}$, the more particles are created (Fig.~\ref{fig:pqcd}).
The average particle energy decreases as they propagate outward, until
it reaches $\overline{E}_f \sim 300$ MeV at the photosphere edge, where
hadronization occurs.

\begin{figure}[h!]
\includegraphics[width=0.4\textwidth]{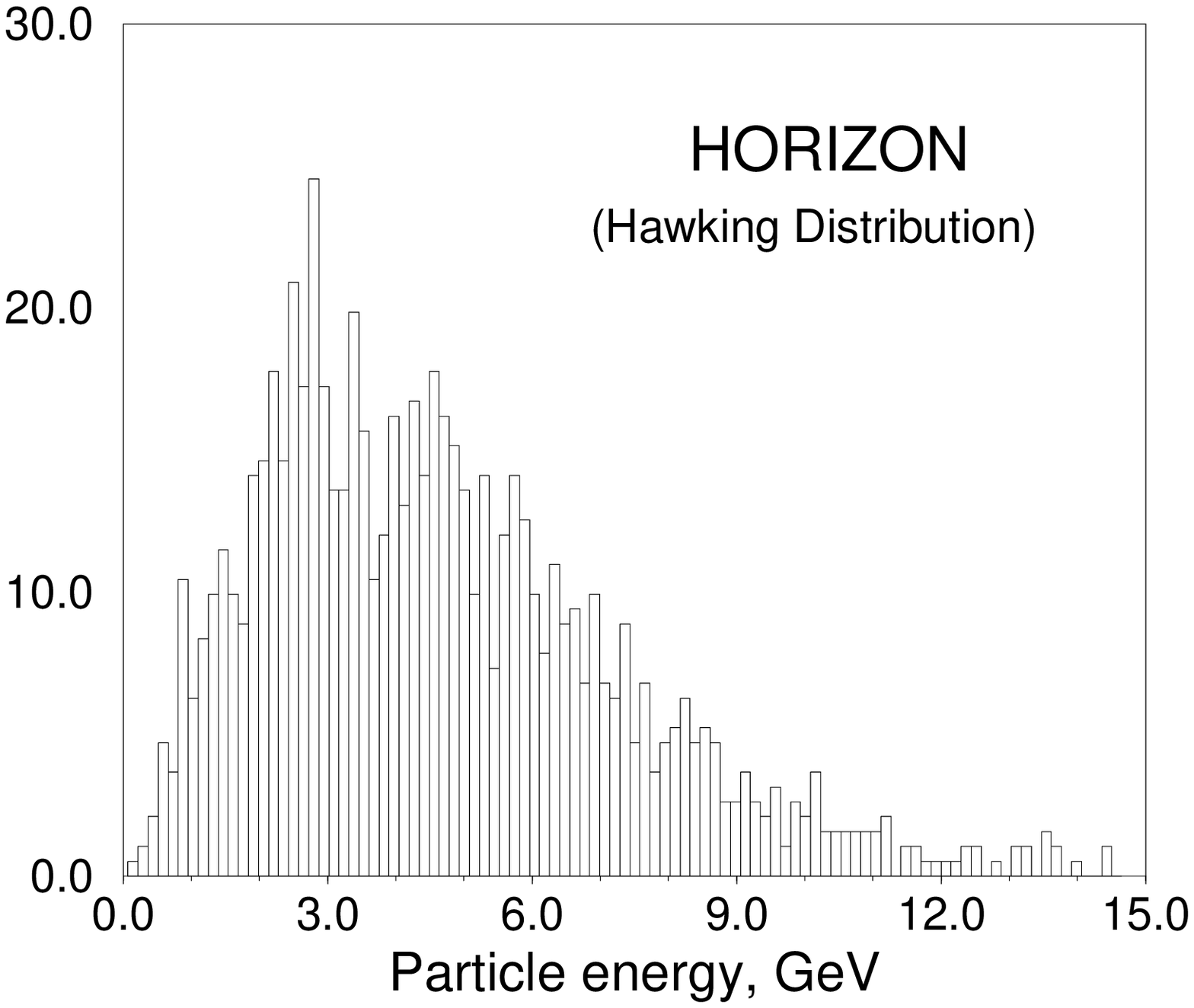}
\includegraphics[width=0.4\textwidth]{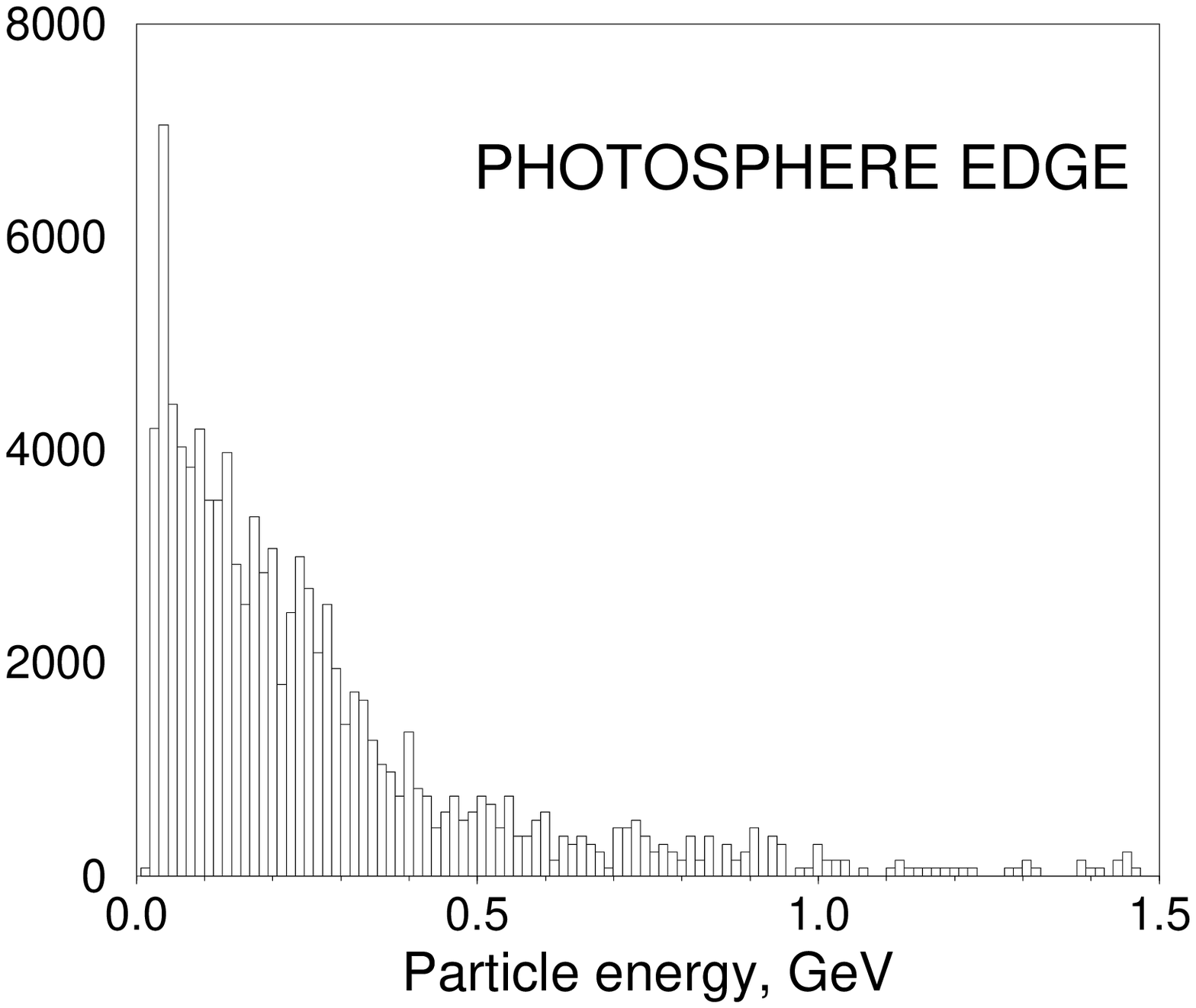} 
\caption{Energy
distribution of quarks and gluons at the horizon of a 1.5 GeV black
hole and at the edge of the QCD photosphere} 
\label{fig:QCDspectra}
\end{figure}

The results for the photosphere parameters for several characteristic
black hole temperatures are summarized in Table~\ref{tab:QCD}.
\begin{table*}[t!] 
\centering
\begin{tabular}{|r|c|c|c|c|} \hline
 & & & & \\ $T_{BH}$ & $r_i\sim r_h$ & $r_f$ & $\overline{E}_i$ & $\Ef$
\\ \hline
 & & & & \\ 200 MeV  & $2.2 \mbox{ fm}$ & $4.9\mbox{ fm}$
	   & $600$ MeV & $300$ MeV\\
 & & & & \\ 1 GeV  & $0.4\mbox{ fm}$ & $19\mbox{ fm}$
			     & 3.0 GeV & 300 MeV \\
 & & & & \\ 50 GeV & $0.008\mbox{ fm}$ & $49 \mbox{ fm}$
& 156 GeV & 300 MeV \\
 & & & & \\ \hline \end{tabular} 
\caption{QCD photosphere
parameters for several black hole temperatures obtained from the test
particle simulation.} 
\label{tab:QCD} 
\end{table*} 
As is evident, the average particle energy can decrease by several
orders of magnitude in a QCD photosphere. How the full spectrum
changes is illustrated in fig.~\ref{fig:QCDspectra}, where we show the
energy distributions of the particles at the horizon and at the
photosphere edge of a 1.5 GeV black hole.  In this example the
average energy decreases by a factor of 13, and the number
of particles increases by the same factor, given by 
$P(1.5$~GeV$)$ [eq.~(\ref{eq:ppQCD})].
\protect 

Of course it is not the quarks and gluons that might reach a distant
observer, but rather the subsequently produced hadrons and their decay
products, especially the photons.  We will therefore consider the
processes by which the QCD plasma creates a potentially
observable gamma ray or antiproton signal. 

\section{Possible Experimental Con\-se\-quences of the Photo\-sphere}

Finally we would like to see what difference the photosphere makes for
observational cosmology or astrophysics.  From the foregoing it is
clear that the spectrum of radiation from an individual black hole near
the end of its existence will be considerably softened from the usual
expectation based on the Hawking flux, eq.~(\ref{eqn:K1}).  In
addition, the integrated contributions of black holes to the diffuse
gamma ray or cosmic antiproton backgrounds should be shifted to lower
energies.  The first step is to compute how quarks and gluons in the
QCD photosphere fragment into observable particles.  We will then
integrate the individual BH fluxes over time and over the initial mass
distribution of BH's to arrive at the diffuse background fluxes.

\subsection{Hadronization and Final Spectrum of QCD photosphere}

Roughly speaking, the QCD interaction is perturbative ($\alpha_s < 1$)
when the distance between the particles is smaller than
$\Lambda_{QCD}^{-1}$.  This condition is satisfied in the photosphere
region. At larger distances, however, vacuum fragmentation of quarks
and gluons will occur, which is what happens at the photosphere edge.
For an accurate calculation of the final photon spectrum, 
we should first compute the neutral pion flux
coming from the photosphere parton distributions 
using a jet fragmentation code, and then the photon flux resulting
from $\pi^0\to\gamma\gamma$ decays.  
However, following ref.~\cite{Heck:2} we can
estimate this spectrum as a convolution of the quark-gluon spectrum,
available from our test particle simulation, with the pion
fragmentation function \cite{MacWeb,Halzen} and the
Lorentz-transformed spectrum of photons from $\pi^0$ decay:\footnote{We
differ with \cite{Heck:2} concerning the limits of integration here.
The minimum and maximum photon energies from a decaying pion
boosted to energy $E_\pi = \gamma m_\pi$ are $E_\gamma = E_\pi (1 \pm
\sqrt{1-\gamma^{-2}})/2$.  Inverting this equation implies that a
photon of energy $E_\gamma$ can come from pions with energies
satisfying $E_\pi \ge E_\gamma + m_\pi^2/4 E_\gamma$.}
\be
\frac{d^{\,2} N_\gamma}{dE_\gamma\,dt}=\int_{E_0}^\infty \! dE_\pi \ 
\frac{dg_{\pi \gamma}(E_\pi)}{dE_\gamma} \frac{d^{\,2} N_\pi}{dE_\pi\,dt},
\label{eq:start}
\ee
where $E_0=E_\gamma + m_\pi^2/4E_\gamma$. The number of photons of energy 
$E_\gamma$ created by a pion moving with velocity $\beta$ 
and decaying isotropically in its rest 
frame is
\be
\frac{dg_{\pi \gamma}(E_\pi)}{dE_\gamma}=\frac{2}{\gamma m_\pi \beta}
=\frac{2}{\sqrt{E_\pi^2-m_\pi^2}},
\ee
where $\gamma=(1-\beta^2)^{-1/2}$.
The pion flux is \cite{MacWeb}
\be
\frac{d^{\,2} N_\pi}{dE_\pi\,dt}=\sum_j \int_{E_\pi}^\infty \! dQ \ 
\frac{dg_{j \pi}(Q,E_\pi)}{dE_\pi} \frac{d^{\,2}N_j}{dQ\,dt} \ ,
\ee
where the sum is over relevant species in the plasma (quarks and gluons).
For the number of pions with energy $E_\pi$  produced 
per unit energy by each quark or gluon, we use the empirical
fragmentation function \cite{HSW,Halzen}
\be
\frac{dg_{j \pi}(Q,E_\pi)}{dE_\pi}=\frac{15}{16} 
\sqrt{\frac{Q}{E_\pi^3}}
\left( 1-\frac{E_\pi}{Q} \right)^2.
\label{eq:jpi}
\ee
Lastly, $d^{\,2}N_j/dQ dt$ is the quark or gluon flux at the
outer edge of the photosphere.  In what follows we treat quark and
gluon jets equally and thus write $\sum_j d^{\,2}N_j/dQ dt = 
d\dot{N}/dQ$.

Combining (\ref{eq:start}) through (\ref{eq:jpi}) we obtain:
\begin{eqnarray}
\label{eq:convolution}
\frac{d^{\,2} N_\gamma}{dE_\gamma\,dt}&=&\int_{E_0}^\infty \! dE_\pi \ 
\frac{15/8}{E_\pi^{3/2}\sqrt{E_\pi^2-m_\pi^2}}\times  \nonumber \\
 && \!\!\!\!\!\!\!\!\!\!\!\int_{E_\pi}^\infty \! dQ\ 
 \sqrt{Q} 
\left( 1-\frac{E_\pi}{Q} \right)^2 \frac{d\dot{N}}{dQ},
\end{eqnarray}
where $E_0=E_\gamma+m_\pi^2/4E_\gamma$.

\begin{figure}[h!]
\centering
\includegraphics[width=0.5\textwidth]{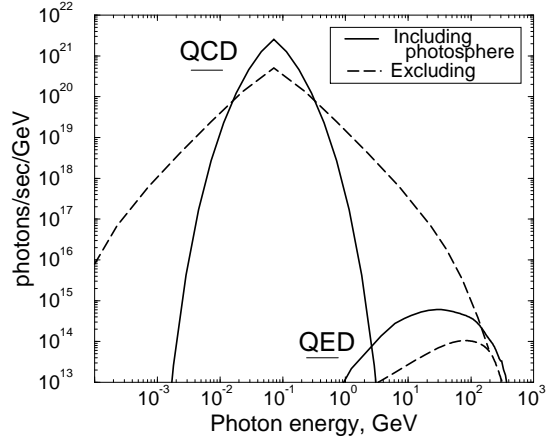}
\caption{Photon emission spectrum from $T=50$ GeV ($M=2\times 10^{12}$g)
black hole. Solid lines are spectra which include photospheres. Dashed 
lines are given for comparison and represent the results for photons 
from direct quark fragmentation at the horizon and subsequent $\pi^0$ decay 
(QCD), and for direct photon emission neglecting the QED photosphere (QED).} 
\label{fig:convolution}
\end{figure}

We have calculated the integral~(\ref{eq:convolution}) numerically for
a large range of black hole temperatures. The results for one of them
($T_{BH}=50$ GeV) are presented in Fig.~\ref{fig:convolution} and
compared to the results obtained neglecting the photosphere, but taking
into account direct quark fragmentation at the horizon and subsequent
$\pi^0$ decay, as in~ref.~\cite{MacWeb}. Also shown are the spectra of
photons emitted directly from the black hole (neglecting the QED
photosphere) and from the QED photosphere, which just starts to form at
this temperature. The actual full spectrum of a 50~GeV black hole is
the addition of the two solid lines.  The results are in agreement with
fig.~1 of ref.~\cite{Heck:2}, except for the QED photosphere spectrum, as 
discussed above.
\protect

\protect
\begin{figure}[h!]
\centering
\includegraphics[width=0.5\textwidth]{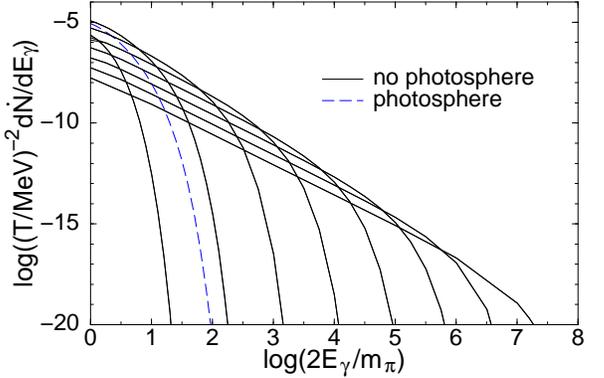}
\caption{Photon emission spectrum from $\pi^0$ decay for black holes,
ignoring photosphere, with $T_{BH} = 50$ MeV (narrowest), increasing 
$T_{BH}$ by factors of 10, up to 500 TeV (widest).  Dashed curve shows
the spectrum including photosphere (see text for normalization).}
\label{fig:gamma_pi_spect}
\end{figure}

We see that the photon flux from the QCD photosphere peaks at
an energy of $m_\pi/2$; if the pions were at rest then all photons 
would have this energy, but $E_\gamma$ is Doppler shifted by the
pion motion, and the width of the distribution grows with the average
energy of the pion.  Figure \ref{fig:gamma_pi_spect} shows the photon
spectra for a large range of BH temperatures in the case where the
photosphere is neglected.  As a function of $y\equiv
\log(2 E_\gamma/m_\pi)$, these spectra are symmetric under $y\to -y$.
Because the total power output of the BH goes like $\int dQ\, Q
d\dot{N}/dQ\sim T_{BH}^2$, it is convenient to display instead the
quantity $T_{BH}^{-2} d\dot{N}_\gamma/dE_\gamma$, as we have done 
in Fig.\ \ref{fig:gamma_pi_spect}.  We fit these functions to 10th order
polynomials,
\begin{eqnarray}
	&&\log_{10}(T_{BH}^{-2} d\dot{N}_\gamma/dE_\gamma) 
	\cong \sum_i a_n |y|^n;	\nonumber\\
	&& y \equiv \log_{10}
	2 E_\gamma/m_\pi	
\label{powerser}
\end{eqnarray}
whose coefficients (up to 8th order, which is still a good
approximation) are given in table \ref{tab:coeff}.  The absolute
normalization follows from the ansatz $d\dot{N}/dQ = (Q/T_{BH})^2
\exp(-Q/T_{BH})/(3\,!)$ which we took as an approximation to the actual
Hawking distribution (up to an overall normalization factor which can
be computed) to obtain these functions.  They will be useful later when
we integrate the BH radiation over time.   Although the spectra are
really analytic at $y=0$, the derivative changes so quickly there that
the best fit is obtained by using an odd function for $y>0$ and then
letting $y\to |y|$ to cover the $y<0$ side.  Also shown in
fig.~\ref{fig:gamma_pi_spect} is the corresponding spectrum, divided by
$T_0^2$, for the photosphere, whose parton flux is taken to be
$d\dot{N}/dQ = \exp(-Q/T_{0})$, where $T_0 = 300 $ MeV.  Like the
nonphotosphere results, a photon spectrum computed using this
distribution should be multiplied by a factor of $T_{BH}^2$ (for
$T_{BH}>T_0$) to represent a BH with horizon temperature $T_{BH}$, to
insure that the power output at the photosphere is equal to that at the
horizon for any value of $T_{BH}$.
\begin{table*}[!t]
\centering
\begin{tabular}{|r|r|r|r|r|r|r|r|r|r|}
\hline
$T_{BH}\ \ \ $ & $a_0\ \ $ & $a_1\ \ $ & $a_2\ \ $ & $a_3\ \ $ & $a_4\ \ $ 
& $a_5\ \ $ & $a_6\ \ $ & $a_7\ \ $ & $a_8\ \ $ \\
\hline
50 MeV &$-5.660$ &$-1.634$ & $-3.055$ & $-1.853$ & $1.308$ & $-1.518$ &
$-1.207$ &
$2.169$ & $-1.596$ \\
\hline
500 MeV & $-4.938$ & $-0.973$ & $-1.059$ & $-0.353$ & $1.544$ & $-2.191$
& $1.566$ & $-0.710$ & $0.189$ \\
\hline
5 GeV & $ -5.289$ & $-0.822$ & $ -1.124$ & $ 1.016$ & $ -0.818$ & $ 0.6191$
& $ -0.418$ & $ 0.195$ & $ -0.057$ \\
\hline
50 GeV & $ -5.882$ & $0.237$ & $-4.838$ & $7.965$ & $-8.251$ & $5.528$
& $ -2.430$ & $ 0.695$ & $ -0.125$ \\
\hline
500 GeV & $ -6.270$ & $ -0.795$ & $ -1.136$ & $ 1.105$ & $ -0.747$
& $ 0.378$ & $ -0.146$ & $ 0.041$ & $ -0.008$ \\
\hline
5 TeV & $-6.770$ & $-0.777$ & $-1.243$ & $ 1.338$ & $ -0.991$ & $ 0.514$
& $-0.184$ & $ 0.044$ & $ -0.007$ \\
\hline
50 TeV & $ -7.270$ & $ -0.735$ & $ -1.402$ & $ 1.549$ & $ -1.105$
& $ 0.521$ & $ -0.163$ & $ 0.033$ & $ -0.004$ \\
\hline
500 TeV & $ -7.770$ & $ 7.316$ & $ -24.23$ & $ 27.72$ & $ -17.30$
& $ 6.561$ & $ -1.580$ & $ 0.244$ & $ -0.023$ \\
\hline
photosphere & $ -5.070$ & $ -1.177$ & $-1.629$ & $ -0.515$ & $ 1.506$
& $ -1.965$ & $ 1.155$ & $ -0.411$ & $ 0.0648$ \\
\hline
\end{tabular}
\caption{Coefficients for photon fluxes from $\pi^0\to\gamma\gamma$ used
in eq.~(\ref{powerser}).}
\label{tab:coeff}
\end{table*}

\subsection{Diffuse gamma ray background}

Finding the contribution of black hole radiation to the diffuse
photon background consists of two steps: (1) first integrate the
contribution of a single BH over time, and (2) integrate this result 
over the initial mass distribution of the individual BH's.  Let us
denote the time-integrated spectrum emitted by a single black hole
by $dN_1/dE_\gamma$.  Going from an initial time $t_i$ to the final time
$t_f$, and accounting for the redshifting of the photons between the
time of emission and the present ($t_0$),
\be
	{dN_1\over dE_\gamma} = \int_{t_i}^{t_f} dt\, Z(t)
	{d\dot{N}_\gamma\over dE_\gamma}
	[Z(t) E],
\ee
where $d\dot{N}_\gamma/dE_\gamma$ 
is the $\pi^0\to\gamma\gamma$ flux derived
in the previous section.
Notice that we must multiply both $d\dot{N}_\gamma/dE_\gamma$ 
and its argument by the
redshift factor $Z\equiv (1+z)$.  It is more convenient to integrate
over the BH temperature, however.  The time-temperature relation can 
be found by equating the rate of change of the BH mass with its power 
output, to obtain \cite{Page,Halzen}
\begin{eqnarray}
	{dT\over dt} &=& \bar{\alpha}(T)\, G\, T^4,\nonumber\\
	\bar{\alpha} &=& 0.57\, d_{1/2} + 0.23\, d_{1},
\label{timetemp}
\end{eqnarray}
where $d_s$ is the number of degrees of freedom (spin, charge and color)
of spin $s$ which can be emitted by the BH at the given temperature.
If we ignore the weak $T$-dependence of $\bar\alpha$, eq.~(\ref{timetemp})
gives the time-temperature relation 
\be
	t-t_i = {1\over 3\bar\alpha}(T_i^{-3} - T^{-3})
\ee
for a BH with initial temperature $T_i$ at time $t_i$.  Let us now
assume that $t_i = 0$ and define $T_*$ as the initial temperature of
a black hole which is disappearing today, $T_*\approx 100$ MeV.
Then $t_0 = 1/(3\bar\alpha T_*^3)$, and we can write the redshift factor
as
\be
	Z = 1+z = \left({t_0\over t}\right)^{2/3} = \left(
	\left({T_*\over T_i}\right)^3 - \left({T_*\over T}\right)^3
	\right)^{-2/3}.
\ee
Thus 
\be
	{dN_1\over dE_\gamma} = \int_{T_i}^{T_f(T_i)} 
	\!{dT\over \bar\alpha
	\, G\, T^4} 
	Z(T){d\dot{N}_\gamma\over dE_\gamma}[Z(T)E_\gamma].
\label{intT}
\ee
The final temperature in this expression is given by
\be
	T_f = \left\{ \begin{array}{cc} 
	T_i (1-(T_i/T_*)^3)^{-1/3}, & T_i < T_*;\\
	\infty,			    & T_i \ge T_*.
	\end{array}\right.
\ee

Next we must add up the contributions from all black holes.  The 
distribution of initial BH masses is taken to be \cite{Mac:Carr}
\be
	{dN\over dM_i} = {(\beta-2)\Omega_{BH}\rho_c\over M_*^2}
	\left({M_i\over M_*}\right)^{-\beta} \equiv C_M M_i^{-\beta},
\ee
where $\Omega_{BH}$ is the fraction of the critical density of the
universe which is in primordial black holes, 
$\beta = 2.5$ for the usual equation of state $p = \rho/3$,
and $M_* \approx 10^{15}$ g is the mass of a BH with $T_i=T_*$ (hence a
BH which is disappearing in the present epoch).  
Then, trading $M_i$ for $T_i$, the integral over initial BH's gives
the spectrum of diffuse gamma rays as
\be
  {dN_\gamma\over dE_\gamma} = C_M \int_0^\infty dT_i\, T_i^{\beta-2}
  \,{dN_1\over dE_\gamma}.
\label{intTi}
\ee

\begin{figure}[h!]
\centering
\includegraphics[width=0.45\textwidth]{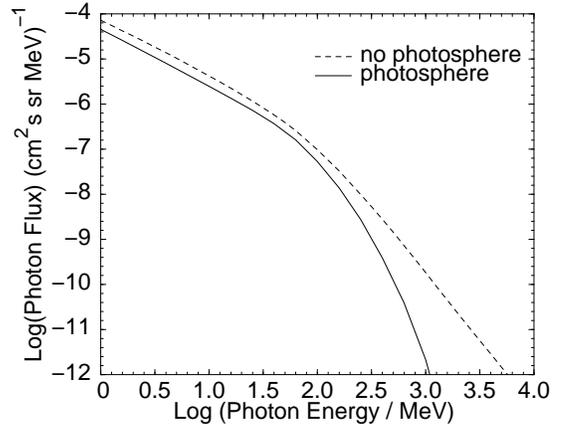}
\caption{Theoretical prediction for the diffuse gamma ray spectrum
from decaying black holes, with and without the QCD photosphere,
assuming $\Omega_{BH} = 7.6\times 10^{-9} h_0^{-2}$.}
\label{fig:int_spect}
\end{figure}

The photon flux per unit energy is $dN_\gamma/dE_\gamma$ times the
speed of light and a geometric factor, $(4\pi)^{-1}\int_0^{2\pi} d\phi
\int_0^1 d(\cos\theta) \cos\theta = 1/4$.  We have computed this flux
both with and without the QCD photosphere to see the effect of the
latter.  The result is shown in fig.~\ref{fig:int_spect}, where we have
normalized the no-photosphere curve to agree with the predictions of
ref.~\cite{Mac:Carr} at $E=100$ MeV, for the case of $\Omega_{BH} =
7.6\times 10^{-9} h_0^{-2}$, which saturates the experimental limit.
One sees that although the photosphere dramatically suppresses the
spectrum for $E> 100$ MeV, the effect is small at lower energies.
\protect

\begin{figure}[h!]
\centering
\includegraphics[width=0.5\textwidth]{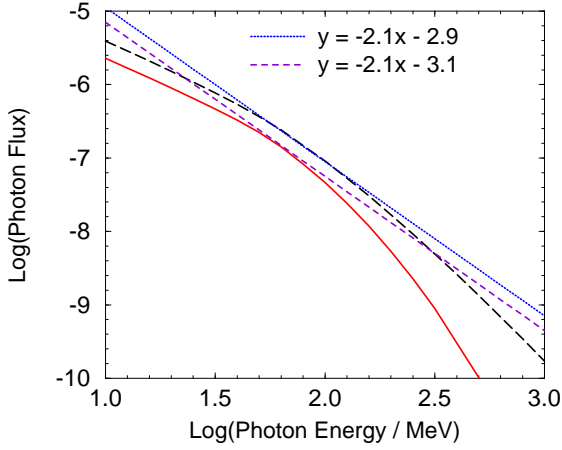}
\caption{Close-up of the 100 MeV region in the previous figure,
showing where a line with the slope of the observed spectrum becomes
tangent to each curve.}
\label{fig:int_zoom}
\end{figure}

The 100 MeV energy range is the most important one for setting limits on
the primordial BH contribution to the energy density of the universe,
for the following reason.  The theoretical prediction for the diffuse
background spectrum goes like $E^{-1}$ at low energies and $E^{-3}$ at
high energies, with $E\sim 100$ MeV $\sim m_\pi$ being the region where
the slope changes.  On the other hand, the extragalactic flux measured 
by the Energetic Gamma Ray Experiment Telescope (EGRET) has an $E^{-2.1
\pm 0.03}$ energy spectrum \cite{EGRET},
 intermediate between the two theoretical
slopes.  Therefore as one increases $\Omega_{BH}$ from small values,
the first place where the theoretical prediction comes into conflict
with the observation is at the ``knee'' of the theoretical spectrum.
The photosphere has only a 60\% effect on the flux at these energies.
In figure \ref{fig:int_zoom} we show where a line with the observed
slope first encounters the predicted spectra as $\Omega_{BH}$ is
increased.  The intercept decreases by a factor of $10^{0.2} = 1.6$
when the photosphere is taken into account; thus the bound on 
$\Omega_{BH}$ is only slightly weakened.
\protect

\subsection{Antiproton background}
A similar effect of the photosphere can be found in the predicted flux
of antiprotons, which is interesting because of two current experiments
which search for antimatter coming from cosmic sources, BESS
\cite{BESS} and AMS \cite{AMS}.  Since the effective temperature of the
BH is limited to $\sim 300$ MeV by the QCD photosphere, we expect a
reduction in the flux of protons and antiprotons relative to
predictions using the Hawking spectrum.

The computation is somewhat simpler than for photons since we only need
the fragmentation of quark and gluon jets into antiprotons, with no
additional subequent decay as in the case of $\pi^0\to\gamma\gamma$.  A
rough fit to the fragmentation function can be inferred from actual
data for $e^+ e^- \to $ jets $\to$ hadrons.  Let $Q$ be the quark or
gluon energy, $p_p$ the antiproton momentum, and $x_p = p/Q$  the
momentum fraction.  From fig.~17 of ref.~\cite{TASSO} we find that
the normalized cross section for $p,\bar p$ production can be fit by
\be
	{x_p\over\sigma_{tot}}{d\sigma\over dx_p} \cong
	\left\{\begin{array}{ll} m_1\, \ln x_p + b_1, &
	x_p > x_{max};\\  m_2\, \ln x_p + b_2, &
	x_p < x_{max};\end{array}\right.
\label{linfit}
\ee
where
\begin{eqnarray}
  && m_1 = -0.259, \quad  b_1 = 0.014;\\
  && m_2 = \phantom{-}0.318,\quad b_2 = (m_2-m_1)\ln x_{max} + b_1,
\nonumber
\end{eqnarray}
and the momentum fraction where the distribution peaks, $x_{max}$,
empirically depends on the parton energy according to
\be
	\ln x_{max}  = -0.97 \log(Q/{\rm GeV}) - 0.95.
\ee
Eq.~(\ref{linfit}) only gives a good approximation for values
such that $(x_p/\sigma_{tot})d\sigma/dx_p \gsim 0.03$; the tails
of the distribution are better represented by a gaussian, 
\be
	{x_p\over\sigma_{tot}}{d\sigma\over dx_p} \cong
	F(Q) \exp\left( -{\ln^2 {x_{max}/ x_p}\over \sigma^2(Q)}\right),
\label{tail}
\ee
where the width is supposed to depend on $Q$ like
$\sigma^2(Q) = C(\ln^{3/2} 4Q^2/\Lambda^2 - \ln^{3/2} \mu^2/\Lambda^2)$,
with $\mu\cong 0.35$ GeV, $\Lambda = \Lambda_{QCD}\cong 0.2$ GeV,
$F(Q)$ chosen so that $\int_0^1 d x_p\,(x_p/\sigma_{tot})d\sigma/dx_p  = 1$
and $C$ being a constant.  However we did not find this to be a good 
representation of the actual data in the vicinity of the peak,
for any constant value of $C$.  We have thus relegated the form 
(\ref{tail}) for representing the tails of the distribution, which
in any case make a subdominant contribution to the final antiproton flux.

\begin{figure}[h!]
\centering
\includegraphics[width=0.5\textwidth]{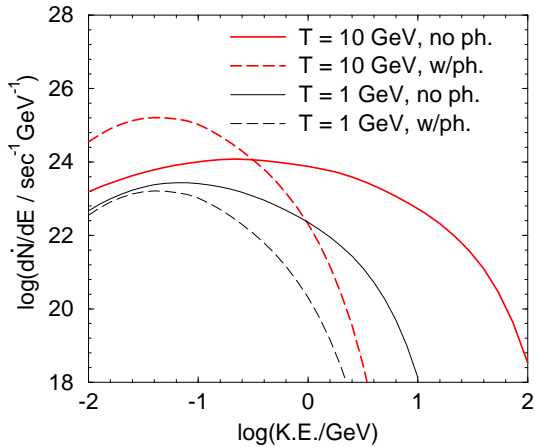}
\caption{Instantaneous $p,\bar p$ fluxes from BH's with 
$T_{BH} = 1$ GeV (solid lines) and 10 GeV (dashed lines), including 
(``w/ph.'') or neglecting (``no.ph.'') the QCD photosphere, plotted
as a function of the kinetic energy.}
\label{fig:pbar}
\end{figure}

The fragmentation function is related to the cross section by
\be
	{dg\over dE_p} = {1\over\sigma_{tot}}{d\sigma\over dE_p}
	= {E\over p^2}\, {x_p\over\sigma_{tot}}{d\sigma\over dx_p}.
\ee
The $p,\bar p$ flux from a single black hole is then
\be
   {d\dot{N_p}\over dE_p} = \int_{E_p}^\infty dQ\,{dg\over dE_p}(Q,E_p)
   {d\dot{N}\over dQ}.
\ee
We show the numerical results for four different cases in fig.\
\ref{fig:pbar}: black holes with $T = 1$ GeV and $T=10$ GeV, ignoring
the photosphere, and the same BH's taking the photosphere into
account.  We have used the results shown in fig.~1 of ref.\ 
\cite{Mac:Carr}
to fix the absolute normalization. The only effect of the horizon
temperature, when the photosphere effects are included, is to multiply
the distribution by a factor of $(T/T_0)^2$ for $T>T_0$, where $T_0 =
300$ MeV is the effective temperature of the photosphere.  This factor
comes from demanding that the total power output of the BH is the same
with or without the photosphere.
\protect

\begin{figure}[h!]
\centering
\includegraphics[width=0.5\textwidth]{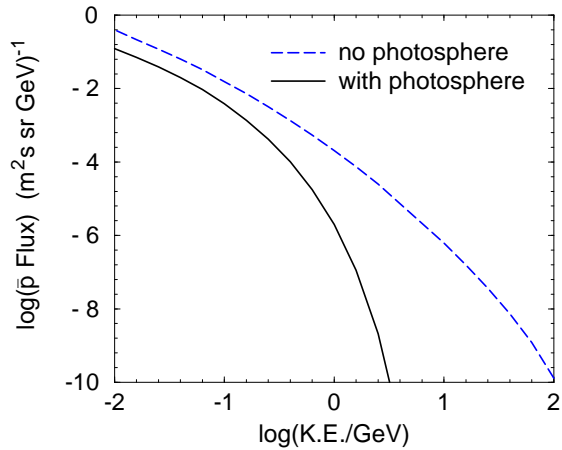}
\caption{Diffuse background $p,\bar p$ flux from integrating over
time and initial distribution of BH's, as a function of the 
kinetic energy. 
 $\Omega_{BH}$ is the same as in fig.\ \ref{fig:int_spect}.}
\label{fig:pbar_int}
\end{figure}

To find the diffuse antiproton flux, we should integrate over
the black hole temperature and the distribution of initial masses
(temperatures) as we did for photons, eqs.~(\ref{intT}) and 
(\ref{intTi}).  The only difference is that we are interested in
nonrelativistic as well as relativistic protons, so we must redshift
the momentum rather than the energy.  Instead of 
the factor $Z d\dot{N}_\gamma/dE\gamma [ZE\gamma]$ in eq.~(\ref{intT}), we get
\be
	Z {d\dot{N}_p\over dE_p} [ZE_p] \to {Z^2 E_p\over E_0}
	 {d\dot{N}_p\over dE_p}[E_0]
\ee
where $E_0 = \sqrt{Z^2(E_p^2-m^2) + m^2}$.  The result (normalized to
agree with fig.\ 2 of ref.~\cite{Mac:Carr}) is shown in fig.\
\ref{fig:pbar_int}.  Again, we take the BH density to be the maximum
allowed by the gamma ray background, 
$\Omega_{BH} = 7.6\times 10^{-9} h_0^{-2}$.
\protect

The recent observations by BESS give an antiproton flux of $(8\pm
2)\times 10^{-3}$m$^{-2}$ sr$^{-1}$sec$^{-1}$GeV$^{-1}$ at $E_p =
0.2-0.3$ GeV, and higher values up to $2\times 10^{-2}$ at larger
energies; thus the bound on $\Omega_{BH}$ from antiprotons is set by
the low energy range of the observations.  Comparison with fig.\
\ref{fig:int_spect} shows that the nonphotosphere prediction for the
$\bar p$ flux is somewhat in excess of the data at these energies,
suggesting that one might be able to set slightly stronger limits on
$\Omega_{BH}$ using the antiproton flux rather than gamma rays.
However, the predicted flux taking into account the photosphere is
$10^{0.6}\cong 4$ times weaker, and in better agreement with the
data.  

It might seem surprising that the low energy $\bar p$ flux is not
degraded more than it is by the photosphere at kinetic energies below
$m_p$.  Apparently it is the tails of the distributions of the underlying
partons which are mostly responsible for producing low-energy protons,
and the largest contribution will come from black holes with temperatures
somewhat below the nucleon mass.  In this regime the difference between
having the QCD photosphere or not is minimized.

\section{Conclusion}

Our main results can be summarized as follows. 
The test-particle method of solving the Boltzmann equation, previously 
used for analyzing heavy ion collisions, was applied to the problem of 
black hole evaporation. The method was adapted to the situation of 
steady-state diffusion of particles emitted by a microscopic black 
hole. A code to simulate the bremsstrahlung and pair production 
interactions of the test particles was developed, leading to solutions 
for the particle distribution functions at any distance from the black 
hole horizon.

Simulation of microscopic black hole emission in both QED and QCD
energy ranges corroborates the idea of photosphere formation pioneered
by Heckler in \cite{Heck:1}. We find that any black hole of mass $M \le
5 \times 10^{14}$ g develops a cloud of interacting quarks and gluons
which extends a certain distance from the black hole horizon. The
evolution of such small ($r < 0.08$ fm) black holes is dominated by
mass loss through Hawking radiation. Part of this radiation is in free
quarks and gluons which are processed in the QCD photosphere until
their average energy drops to the point $\overline{E} \sim
\Lambda_{QCD}$, where they hadronize into stable particles and
fast-decaying pions. Another part consists of electrons, positrons and
photons. Once the black hole mass drops below $M \sim 2\times 10^{12}$
g, these particles interact significantly enough to form another, less
dense cloud at a distance about 700 times the horizon radius. This QED
photosphere extends over a distance of about 400 fm, where it
dissipates and emits much less energetic, but more numerous, electrons,
positrons and photons.

Energy distributions of the particles leaving both photospheres were
obtained and shown to greatly differ from the original nearly-thermal
Hawking distributions by being softened to much lower average energies:
$\bar E \sim 300$ MeV for QCD,  $\bar E$ ranging from 100 GeV for a
$10^{12}$ g black hole $\sim 0.5$ MeV for a $10^{6}$ g black hole in
the QED case.  We used the QCD spectra to compute the contributions
of individual black holes and all BH's in the universe to potentially
observable gamma ray and antiproton signals, and compared to the
previous expectations based on ignoring the photosphere.  In the 
regions where the experimental sensitivity is greatest, the photosphere
lowers the fluxes by only a small factor:  1.6 for photons and 4 for
antiprotons.

Our findings do not support the approximation made in
ref.\ \cite{Heck:1} of treating the photosphere as a fluid.  Rather we
get a picture of a steadily expanding cloud of particles which never
quite thermalizes, and has interactions which are primarily low in
momentum transfer.  This is how we interpret the fact that our
Boltzmann code gives much smaller photosphere, hence much less energy
degradation of particles, in the case of the QED photosphere, than
claimed in \cite{Heck:1}.  This discrepancy did not appear for the QCD
case because there both approaches find that the photosphere ends when
hadronization begins.  Furthermore we do not find a relativistically
expanding photosphere, which was also claimed in \cite{Heck:1} on the
basis of the fluid approach.  This led us to find larger differences
in the diffuse gamma ray background between the photosphere and
nonphotosphere predictions than found by ref.~\cite{Heck:2}.  The
reason is that \cite{Heck:2} undoes the energy-degrading effects
of the photosphere to a large extent by boosting the distributions
from the fluid frame to the observer frame, a step which is not
necessary in our approach, since we always work in the latter 
reference frame.

It is disappointing that the observable consequences of the photosphere
are small in the experimentally interesting energy ranges.  If we
were lucky enough to have a nearby BH reach the end of its existence
however, a real experimental test might be possible, since the 
spectra for individual BH's have radically different characteristics
with or without a photosphere.  On the theoretical side, the question
of whether black hole photospheres indeed develop hinges crucially 
on whether the range of the \br\ interaction is really $1/m_e$
or something effectively shorter.  Although the arguments supporting
this claim look plausible, it is perhaps deserving of more detailed
study.

\medskip
{\bf Acknowledgments.}  We wish to thank F.\ Halzen, K.\ Ragan and 
S.\ Orito for discussions or for providing helpful information.
This work was supported by the Natural Sciences and Engineering
Research Council of Canada.




\begin{thebibliography}{2222222}
%
\bibitem{Hawking}
S.W.~Hawking, Commun.\ Math.\ Phys.\ \textbf{43}, 199 (1975).
%
\bibitem{Mac:Carr}
J.H.~MacGibbon and B.J.~Carr, Astrophys.\ J.\ \textbf{371}, 447 (1991).
%
\bibitem{Halzen}
F.~Halzen, E.~Zas, J.H.~MacGibbon and T.C.~Weekes, Nature
\textbf{353}, 807 (1991).
%
\bibitem{Page}
D.N.~Page, Phys.\ Rev.\ \textbf{D13}, 198 (1976).
%
\bibitem{Inflation}
B.J.~Carr, J.H.~Gilbert and J.E.~Lidsey, Phys.\ Rev.\ \textbf{D50}, 4853
	(1994);\\
R.~Bousso and S.W.~Hawking, Phys.\ Rev.\ \textbf{D54}, 6312 (1996);\\
J.S.~Bullock and J.R.~Primack, Phys.\ Rev. \textbf{D55}, 7423 (1997).
%
\bibitem{QCD}
K.~Jedamzik, Phys.\ Rev.\ \textbf{D55}, 5871 (1997);\\
C.Y.~Cardall and G.M.~Fuller, astro-ph/9801103 (1998).
%
\bibitem{Halzen2} F.\ Halzen, B.\ Keszthelyi and E.\ Zas, 
	Phys.\ Rev.\ D52 (1995) 3239.
\bibitem{Heck:1}
A.F.~Heckler, Phys.\ Rev.\ \textbf{D55}, 480 (1997).
%
\bibitem{MacWeb}
MacGibbon, J.H. and Weber, B.R. Phys. Rev. D \textbf{41}, 3052 (1990).
%
\bibitem{Ol:Hill}
J.~Oliensis and C.T.~Hill, Phys.\ Lett.\ \textbf{B143}, 447 (1984).
%
\bibitem{DasGupta}
G.F.~Bertsch and S.~Das Gupta, Phys.\ Rep.\ \textbf{160}, 191 (1988).
%
\bibitem{Welke}
G.M.~Welke, McGill University Ph.D.\ Thesis (1990).
%
\bibitem{JR}
Jauch, J.M. and Rohrlich, F. \emph{The Theory of Electrons and Photons}
(Springer-Verlag, New York, 1975).
%
\bibitem{Haug}
Haug, E. Z. Naturforsch. Teil A \textbf{30A}, 1099 (1975).
%
\bibitem{Weldon}
Weldon, H.A. Phys. Rev. D \textbf{26}, 2789 (1982).
%
\bibitem{RevModernPhysics}
Joseph, J. and Rohrlich, F. Rev. Mod. Phys. \textbf{30}, 354 (1958).
%
\bibitem{Halzen:Martin}
Halzen, F. and Martin, A.D. \emph{Quarks and Leptons}
(John Wiley and Sons, 1984).
%
\bibitem{Barnes}
W.S.~Barnes {\it et al.}, Phys.\ Rev.\ \textbf{117}, 226 (1960).
%
\bibitem{Heck:2}
Heckler, A.F. Phys.\ Rev.\ Lett.\ \textbf{78}, 3430 (1997).
%
\bibitem{LPM}
L.D. Landau and I. Pomeranchuk, Sov. Phys. Dokl. \textbf{92}, 553 
(1953); A.B. Migdal, Phys. Rev. \textbf{103}, 1811 (1956); E.L. Feinberg 
and I. Pomeranchuk, Nuovo Cimento Suppl. \textbf{3}, 652 (1956).
%
\bibitem{HSW} C.T.~Hill, D.N.~Schramm and T.P.~Walker, Phys.\ Rev.\
\textbf{D36}, 1007 (1987).
%
\bibitem{EGRET} EGRET Collaboration, P.~Sreekumar {\it et al.,} astro-ph/9709257, to 
appear in Astrophys./ J./ (1997).
%
\bibitem{BESS} BESS Collaboration (A.\ Moiseev {\it et al.}), Astrophys.\ J.\ 
\textbf{474}, 479 (1997)
%
\bibitem{AMS} R.\ Battiston, Nucl.\ Phys.\ Proc.\ Suppl.\ \textbf{65},
	19 (1998) 
%
\bibitem{TASSO} TASSO Collaboration, M.~Althoff {\it et al.,} Z.\ Phys.\ 
\textbf{C22}, 302 (1984).
%

\end{thebibliography}
\end{document}